\documentclass{aa}
\usepackage{txfonts}
\usepackage{siunitx}
\usepackage{xurl}

\usepackage{bm}

\usepackage{graphicx}   
\usepackage{subcaption}
\usepackage{xcolor}
\newcommand{\bnabla}{\bm{\nabla}}
\newcommand{\bcdot}{\bm{\cdot}}

\newcommand{\rmn}[1]{\mathrm{#1}}

\defcitealias{Perrone2022}{PL22a}
\defcitealias{Perrone2022a}{PL22b}
\defcitealias{Perrone2023}{PBP23}

\usepackage{hyperref}
\hypersetup{
        colorlinks=true,
        breaklinks=true,
        citecolor=blue,
        allcolors=blue
}

\makeatletter
\renewcommand*\aa@pageof{, page \thepage{} of \pageref*{LastPage}}
\makeatother

\newcommand{\bigO}{\mathcal{O}}
\graphicspath{{./Figures/}}

\begin{document}

\title{Thermal conductivity with bells and whistlers: Suppression of the magnetothermal instability in galaxy clusters}
 \titlerunning{Whistler suppression of the magnetothermal instability}

\author{Lorenzo Maria Perrone\inst{1}
                \and
                Thomas Berlok\inst{2,1}
                \and
                Christoph Pfrommer\inst{1}
        }
        
        \institute{Leibniz-Institut f\"{u}r Astrophysik Potsdam (AIP), 
                An der Sternwarte 16, D-14482 Potsdam, Germany\\
                \email{lperrone@aip.de}
      \and
           Niels Bohr Institute, University of Copenhagen, Blegdamsvej 17, 2100 Copenhagen, Denmark
        }

\date{\today}

\abstract{
In the hot and dilute intracluster medium (ICM) in galaxy clusters, kinetic plasma instabilities that are excited at the particle gyroradius may play an important role in the transport of heat and momentum, thus affecting the large-scale evolution of these systems.
In this paper, we continue our investigation of the effect of whistler suppression of thermal conductivity
on the  magneto-thermal instability (MTI), which may be active in the periphery of galaxy clusters and may contribute to the observed levels of turbulence.
We use a subgrid closure for the heat flux inspired from kinetic simulations
and show that MTI turbulence with whistler suppression exhibits a critical transition as the suppression parameter is increased: for modest suppression of the conductivity, the turbulent velocities generated by the MTI decrease accordingly, in agreement with scaling laws found in previous studies of the MTI. However, for suppression above a critical threshold, the MTI loses its ability to maintain equipartition-level magnetic fields through a small-scale dynamo (SSD), and the system enters a ``death-spiral.'' 
We show that analogous levels of suppression of thermal conductivity with a simple model of flat uniform suppression would not inhibit the dynamo.
We propose a model to explain this critical transition, and speculate that conditions in the hot ICM  are 
such that in substantial portions of the galaxy cluster periphery the MTI might struggle to sustain its own dynamo. 
We then look at spatial correlations and energy transfers in spectral space and find that, with whistler suppression, most of the heat is transported along thin bundles of strong magnetic fields (the Autobahns of electrons), while high-$\beta$ regions are brought out of thermal equilibrium. We link this behavior  to the intermittent nature of magnetic fields, and we observe an overall reduction of the efficiency of MTI turbulent driving at the largest turbulent scales.
Finally, we show that external turbulence interferes with the MTI and leads to reduced levels of MTI turbulence. While individually both external turbulence and whistler suppression reduce MTI turbulence, we find that  they can exhibit a complex interplay when acting in conjunction, with external turbulence boosting the whistler-suppressed thermal conductivity and even reviving a ``dead'' MTI. Our study illustrates how extending magnetohydrodynamics with a simple prescription for microscale plasma physics can lead to the formation of a complicated dynamical system and demonstrates that further work is needed in order to bridge the gap between micro- and macro scales in galaxy clusters.
}


\maketitle



\section{Introduction}

In the hot and dilute plasma that fills galaxy clusters known as the intracluster
medium (ICM), transport processes are expected to be significantly modified compared to the standard collisional picture due to the presence of magnetic fields \citep{Sarazin1988}. First, despite the relatively high plasma $\beta$ \citep{Carilli2002,Vogt2005,Kuchar2011}, charged particles gyrate around magnetic field lines on much shorter timescales than Coulomb scattering, and as a result the transport of heat and momentum takes place primarily along the local direction of the magnetic field \citep{Braginskii1965a}. 
This results in a new class of buoyancy instabilities that bypass the stable entropy stratification of galaxy clusters, and that can perturb the ICM respectively in the cluster periphery \citep[the MTI;][]{Balbus2000} or in the core of cool-core clusters \citep[the heat-flux driven buoyancy instability,   HBI;][]{Quataert2008}. The presence of the MTI may have implications for the re-acceleration of volume-filling, relativistic electrons far in the periphery of galaxy clusters \citep{Cuciti2022,Beduzzi2023}.
Second, high-$\beta$, weakly collisional plasmas such as the ICM are a fertile ground for microinstabilities that grow on kinetic scales \citep{Schekochihin2006,Schekochihin2008} and that may substantially impede particle motion. Among these, we mention the mirror, firehose, and whistler instabilities \citep{Chandrasekhar1958,Parker1958,Gary1993,Gary2000}.

A large body of numerical work \citep{Kunz2014,Riquelme2016,Melville2016} and observational evidence from the solar wind \citep[also a dilute and magnetized plasma similar to the ICM;][]{Bale2009,Chen2016} suggests that these microinstabilities act in such a way as to counter the pressure anisotropy and to enhance the scattering of charged particles, leading to a suppression of the effective viscosity and thermal conductivity
\citep{Bale2013,Roberg-Clark2016,Roberg-Clark2018,Komarov2018}.
The implications for the thermodynamic evolution of galaxy clusters are potentially far-reaching, and include suppressed dissipation of turbulence \citep{Zhuravleva2019}, plasma heating \citep{Ley2022}, and explosive magnetic field amplification \citep{Mogavero2014,Zhou2023,Rappaz2023}.
Of particular importance is the impact on thermal conductivity, which may play a role in regulating the radiative losses in cool-core clusters \citep{Narayan2001,Zakamska2003,Kunz2012,Jacob2017a,Jacob2017b}, and whose suppression may explain certain observed features such as cold fronts \citep{Markevitch2007,ZuHone2013,ZuHone2016}. Furthermore, the suppression of thermal conductivity can potentially lead to an impairment of the MTI since the instability relies on fast conduction along magnetic field lines and drives turbulent motions that scale with the value of thermal diffusivity \citep[][hereafter \citetalias{Perrone2022,Perrone2022a}]{Perrone2022,Perrone2022a}.  


Predicting the impact of microscale instabilities and the resulting suppression of heat flux on the MTI is not straightforward, and modeling  self-consistently the saturation of microinstabilities within large-scale simulations is currently unfeasible due to the vast scale separation.
For this reason, several studies have sought to capture the effect of individual microinstabilities through effective models inspired by theoretical considerations and by kinetic plasma simulations. For instance, \citet{Drake2021} suggest through theoretical arguments that whistler suppression of the heat flux could severely constrain the scales where the MTI can efficiently grow in the periphery of galaxy clusters. However, because of the complexity of the physics and the disparate scales involved, these considerations warrant a more detailed investigation and leave the door open to potential surprises.
In the case of the mirror instability, for example, magnetohydrodynamic (MHD) simulations have indicated that the overall impact on the MTI may be comparatively small \citep{Berlok2021} because regions that are mirror-unstable (where conduction is suppressed) coincide with those of fastest linear MTI growth. This demonstrates that the MTI can minimize, to some degree, the onset and impact of deleterious micro-instabilities, and calls into questions simple models where the thermal conductivity is reduced by some fixed uniform factor. 
Finally, the existence of a kinetic counterpart of the MTI that grows on electron kinetic scales (and possibly competes with other high-$\beta$ microinstabilities) further complicates the picture \citep{Xu2016}.


The MTI is not the only source of turbulence in galaxy clusters on large scales. Mergers and (substructure) accretion can stir turbulent motions \citep{Churazov2003,Schuecker2004}, potentially competing with the MTI and suppressing the instability \citep{Sharma2009b,Ruszkowski2010,McCourt2011}, although hints of additive behavior between the MTI and other sources of turbulence have also been reported \citep{Parrish2012}. The interplay between MTI and other sources of turbulence has been previously investigated with local and global numerical setups \citep{McCourt2011,Parrish2012}, which showed that strong external turbulence can dominate over MTI-driven turbulence if the eddy turnover time of the external turbulence is much shorter than the MTI $e$-folding time. In practice, however, given the broad range of scales where the MTI can grow close to its maximum rate this is unlikely to happen at all scales up to the pressure scale-height \citep{McCourt2011}. Some open questions remain, particularly concerning the impact of external forcing on the overall buoyant instability of the ICM with anisotropic thermal conduction \citep{Sharma2009b}.

A number of large-scale cosmological simulations looking at anisotropic heat conduction in galaxy clusters have also been performed \citep{Ruszkowski2010,Ruszkowski2011,ZuHone2013,Yang2016b,Kannan2017,Barnes2019}, but have generally not found strong evidence of MTI action 
against the turbulence produced by cosmological accretion \citep{Ruszkowski2011,Kannan2017,Barnes2019}. 
However, several questions remain on whether these works adequately resolve 
the broad range of scales ($10-100\, \si{kpc}$ in the periphery of galaxy clusters) that are key to energy injection by the MTI, without which the instability would be numerically inhibited. Similarly, high levels of numerical heat diffusion in the low-resolution regions in the outskirts of galaxy clusters will also artificially degrade the growth of the MTI \citepalias{Perrone2022a}.
Carefully constructed large-scale cosmological simulations that adequately resolve the MTI scales are still required in order to draw firm conclusions on the relevance of the MTI in a cosmological context.


In \citet[][hereafter \citetalias{Perrone2023}]{Perrone2023}, we set out to explore the impact of the whistler instability on the MTI through small-scale Boussinesq simulations where we employed a subgrid closure for the heat flux. We found that, depending on the exact parameters of the local environment, the effect on the MTI can range from a mild to a complete suppression of the instability. In particular, we identified the shutdown of the SSD as the culprit behind the death of the MTI in our simulations. We then focused on the interplay between external turbulence and whistler suppression and showed that, with little to no whistler suppression,  externally driven turbulence interferes with the MTI, but does not wipe it out, and that the flow remains overall buoyantly unstable. Contrary to expectations, external turbulence was shown to actually play a beneficial role by reviving the MTI in strongly whistler suppressed simulations. 

In this work we complement the analysis of \citetalias{Perrone2023} along three main scientific drivers: (a) we explain the mechanism behind the critical transition of MTI turbulence with whistler suppression, (b) we explore in what ways whistler suppression is different to a flat uniform suppression of thermal conductivity, and (c) we examine the interaction of whistler-suppressed MTI turbulence with externally driven turbulence.
Our aim is to add another building block to the ongoing conversation of how macro- and micro-turbulence interact in the hot dilute ICM, and to inform future theoretical and numerical work.

This paper is structured as follows. In Sect.~\ref{sec:preliminaries} we briefly review how the presence of magnetic fields and the low collisionality in the ICM affect the standard collisional picture of particle transport, and we introduce our subgrid model for the suppression of thermal diffusivity due to whistler waves. In Sect.~\ref{sec:numerical_methods} we detail our numerical methods, the parameters of the simulations, and the diagnostics employed. In Sect.~\ref{sec:weakening-suppression} we present the first results of our simulations, and focus on the weakening of the MTI with whistler suppression and the appearance of a critical transition above which MTI-driven turbulence dies entirely. In Sect.~\ref{sec:flow-backreaction} we take a closer look at the back-reaction of whistler suppression on the flow and on how the MTI energy fluxes in spectral space are affected. In Sect.~\ref{sec:external-turbulence} we re-examine the role of external turbulence, and show that it can interfere with the workings of the MTI but at the same time be beneficial for simulations where the MTI turbulence is dying.
Finally, we discuss in what respects whistler suppression of thermal diffusivity can be approximated by a flat uniform suppression below the Spitzer value (Sect.~\ref{sec:discussion}) and we draw our conclusions in Sect.~\ref{sec:conclusions}. The main text is complemented by three appendices: in Appendix~\ref{app:chi_derivation} we rewrite the subgrid model of whistler suppression in terms of Boussinesq variables; in Appendix~\ref{app:linear_theory} we show through linear theory how whistler suppression can interrupt the growth of an MTI eigenmode; finally, in Appendix~\ref{app:implementation_whistler_suppression} we detail and validate through benchmarks the numerical implementation of whistler suppression in our code.

\section{Theoretical preliminaries}\label{sec:preliminaries}

\subsection{Collisional transport in the ICM}

In the standard picture of transport processes in plasmas, transport of heat is mediated by particle-particle Coulomb collisions \citep{Spitzer1962}, with electrons (the main particle species responsible for the conduction of heat) scattering off each other every $\lambda_{\text{mfp,e}}$, 
while traveling at the thermal speed $\varv_{\text{th,e}}$. This gives the  Spitzer diffusivity \citep{Spitzer1962}
\begin{align}\label{eq:spitzer-diffusivity}
        \chi_{0} \simeq 4.98 \times 10^{31} \left( \frac{T_\rmn{e}}{5 ~\si{keV}}\right)^{5/2} \left( \frac{n_\rmn{e}}{10^{-3}~\si{cm^{-3}}}\right)^{-1} \si{cm^2.s^{-1}},
\end{align}
(here and in the rest of the paper we express temperature in energy units), which we have normalized for typical values in the periphery of galaxy clusters. However, despite having subthermal magnetic fields \citep[$\beta = p_{\rmn{th}}/ p_{\rmn{mag}} \gtrsim 100$;][]{Carilli2002}, the ICM is strongly magnetized in the sense that the collisional mean-free path of charged particles is much larger than their gyroradius. For   typical conditions in the ICM, the electron mean-free path is
\begin{align}\label{eq:lambda_mfp_spitzer}
        \lambda_{\text{mfp,e}} \simeq 5.83 \left(\frac{n_\rmn{e}}{10^{-3}~ \si{cm^{-3}}} \right)^{-1} \left(\frac{T_\rmn{e}}{5 ~\si{keV}} \right)^2 \si{kpc},
\end{align}
which is many orders of magnitude larger than their gyroradius:
\begin{align}\label{eq:gyroradius_electrons}
        \rho_{\text{e}} \simeq 2.73 \times 10^{-14} \left(\frac{B}{2 ~\si{\mu G}} \right)^{-1} \left(\frac{T_\rmn{e}}{5~ \si{keV}} \right)^{1/2} \si{kpc}.
\end{align}
This means that particles collide against each other very infrequently as they gyrate along magnetic field lines, and transport across the local magnetic field is suppressed. As a result, the heat flux and viscous tensor become anisotropic with respect to the local direction of the magnetic field \citep{Braginskii1965a}. In particular, the anisotropic (or Braginskii) heat flux along the magnetic field takes the form 
\begin{align}\label{eq:braginskii-heat-flux}
    q_{\parallel,\mathrm{s}} \simeq n_\rmn{e} \chi_0 (\bm b \bcdot \bnabla ) T_\rmn{e},
\end{align}
which shows that only temperature gradients along the magnetic field can be thermalized. This property of dilute magnetized plasmas, such as the ICM,
may help explain the presence of features such as cold fronts, which are characterized by sharp temperature gradients on spatial scales comparable or shorter than the electron mean-free path  \citep[see, e.g.,][]{Markevitch2007}. Simulations find that magnetic fields are typically draped around the cold fronts and the suppression of heat conduction across the front, that arises with an anisotropic conduction and this field configuration, has been invoked to explain cold front observations \citep[][]{Lyutikov2006,Asai2007,Dursi2008,Pfrommer2010}, although magnetic draping alone may not be sufficient \citep{ZuHone2013,ZuHone2016}.

\subsection{Whistler suppression of thermal conductivity}\label{sec:model}

The Braginskii form of the heat flux and viscous tensor is derived under the assumption that the plasma is collisional on macroscopic (fluid) scales. In many astrophysical plasmas, including the ICM, this requirement is however only marginally satisfied. As a consequence, a host of kinetic instabilities can grow on small scales (comparable to the particle gyroradius) at a much faster rate than any macroscopic process \citep[see, e.g.,][and references therein]{Schekochihin2008}.
The resulting electromagnetic fluctuations can significantly affect particle trajectories by enhancing their scattering rate, effectively leading to a suppression of the plasma viscosity and thermal conductivity
\citep{Bale2013,Roberg-Clark2016,Roberg-Clark2018,Komarov2018}.
In particular, the whistler instability (which can be excited in high-$\beta$ plasmas by an electron heat flux; \citealt{Levinson1992}, or by an electron temperature anisotropy; \citealt{Gary1996}) has long been identified as a leading mechanism for the suppression of electron transport in weakly collisional plasmas \citep{Levinson1992,Pistinner1998}, and could also appear concomitantly to ion-scale microinstabilities, such as the mirror instability \citep{Ley2023}.
Theoretical arguments and kinetic particle-in-cell simulations showed that at saturation the whistler instability establishes a marginal heat flux which is independent of the temperature gradient and suppressed by a factor of $\sim 1/\beta_{\rmn{e}}$ (where $\beta_\rmn{e}$ is the electron plasma beta) compared to the value it would attain solely due to the streaming of electrons in the absence of electromagnetic fluctuations \citep{Riquelme2016,Roberg-Clark2018,Komarov2018}:
\begin{align}\label{eq:whistler-dominated-hflux}
    q_{\parallel,\mathrm{w}} \simeq 1.5 n_\rmn{e} m_\rmn{e} \varv_{\mathrm{th},\rmn{e}}^3 / \beta_\rmn{e}.
\end{align}
Since in the ICM $\beta \gtrsim 100$, the suppression of the heat flux could then be significant.

As it is currently unfeasible to self-consistently model the saturation of the whistler instability in macroscopic fluid simulations of the MTI, we decided to adopt the following subgrid formula for the heat flux, which is inspired by the work of \citet{Komarov2018} and smoothly interpolates between the collisional heat flux (Eq.~\ref{eq:braginskii-heat-flux}) and the marginal heat flux allowed in the whistler-dominated regime (Eq.~\ref{eq:whistler-dominated-hflux})
\begin{align}
    q_{\parallel} = \left( \frac{1}{q_{\parallel,\mathrm{s}}} + \frac{1}{q_{\parallel,\mathrm{w}}} \right)^{-1} = \frac{q_{\parallel,\rmn{s}}}{1 + \frac{1}{3} \beta_\rmn{e} \lambda_{\text{mfp,e}}/L_{T,\parallel}  }, \label{eq:whistler_suppression_approx}
\end{align}
where $L_{T,\parallel} \equiv \left\lvert \bm b \bcdot \bnabla \ln T_{\text{e}} \right\rvert^{-1}$ is the temperature gradient scale parallel to the magnetic field. Since in the periphery of galaxy clusters the plasma is relatively collisional ($\lambda_{\text{mfp,e}} / L_{T,\parallel} \lesssim 0.1$, see Table~\ref{tab:parameters}), in our model we do not include saturation of the heat flux due to the free-streaming of electrons \citep{Cowie1977}, but depending on the regime of interest other interpolation formulae can be devised \citep[see, e.g.,][]{Beckmann2022}. 

The closure for the heat flux in Equation~\eqref{eq:whistler_suppression_approx} has the advantage of tying together the dynamics on small and large scales, since the large-scale fluid motions determine the degree of whistler suppression through the plasma $\beta$, and are themselves affected by the resulting parallel heat flux. In doing so, it implicitly assumes that, due to the vast separation of temporal and spatial scales, the saturation of the whistler instability is effectively instantaneous with respect to the large-scale MTI motions. While this seems plausible on the macroscopic fluid scales evolved in our simulations, we note that the MTI is not necessarily confined to those scales but exists all the way down to electron kinetic scales, where the whistler instability lives \citep{Xu2016}. At this point, the simple model in Eq.~\eqref{eq:whistler_suppression_approx} is harder to justify, since the two instabilities could interfere with one another. Finally, our model neglects the combined impact of other microscale instabilities, such as the mirror and firehose, which could also change the picture \citep[although individually they are likely subdominant compared to the whistler instability]{Komarov2018}.

\subsection{Theory of MTI turbulence in galaxy clusters' periphery}\label{sec:mti-theory}

In this and the next section we briefly review the theory of MTI turbulence and the Boussinesq approximation. Some readers may choose to skip ahead to Sect.~\ref{sec:estimates_whistler_mti}.

Since the seminal work of \citet{Balbus2000} it was recognized that, in addition to modifying particle transport on small scales, anisotropic thermal conduction can also affect the ICM dynamics on large scales, triggering a new class of buoyancy instabilities. With heat conduction along magnetic field lines, the hydrodynamic stability to convection is determined by the sign of the temperature gradient \citep{Balbus2000}, rather than the entropy gradient \citep{Schwarzschild1906}.  
Under these conditions, two new instabilities arise: the HBI \citep{Quataert2008}, which destabilizes plasmas with a positive temperature gradient with respect to gravity and may be relevant in the central region of cool-core clusters, and the MTI, which can destabilize the periphery of galaxy clusters \citep{Parrish2007,McCourt2011,Kunz2011}, and is the main focus of this paper.  
The MTI operates by extracting energy from the background temperature gradient via conduction of heat along magnetic field lines, which are frozen-in and dragged along with the fluid. As a result, a displaced fluid element remains thermally connected with the plasma at its original location, and to get into pressure balance with its new environment expands (if rising) or contracts (if sinking), becoming buoyantly unstable.

In \citetalias{Perrone2022} and \citetalias{Perrone2022a} it was shown that the MTI produces a state of buoyancy-driven turbulence sustained by energy drawn from the background temperature gradient and composed of density and velocity fluctuations over a wide range of scales, with turbulent properties at saturation that follow clear power laws. In the stably stratified regime typical of the ICM, the rms velocities obey 
\begin{align}
        u_{\text{rms}}^2 \propto \dfrac{\chi \omega_{\rmn{T}}^3}{N^2}, \label{eq:mti-kin-scaling}
\end{align}
where $\omega_{\rmn{T}}$ is the MTI frequency (i.e., the maximum growth rate of the instability), which also corresponds to the typical timescale of MTI turbulence, and $N$ is the Brunt-V\"ais\"al\"a frequency, which determines the buoyant response of the system in a stably stratified environment. Locally (at a given spherical radius $R_0$), they are given by
\begin{align}\label{eq:freq}
        \omega_{\rmn{T}}^2 = -g_0  \left.\frac{\partial \ln T}{\partial R}\right|_0  , \quad
        N^2 = \frac{g_0}{\gamma} \left. \frac{\partial \ln p \rho^{-\gamma}  }{\partial R} \right|_0   , \qquad
\end{align}
where $\rho, p$, and $T$ are the fluid density, pressure, and temperature,
respectively, $g_0 = -(1/\rho) \mathrm{d} p / \mathrm{d} R |_0$ is the local gravitational acceleration (here and in Eq.~\ref{eq:freq} the derivatives are computed at $R_0$), and $\gamma$ is the adiabatic index.
In the periphery of galaxy clusters their values are comparable, and on the order of 
$(600 \,\si{Myr})^{-1} \simeq 5.27\times 10^{-17}\, \si{s^{-1}}$ \citepalias{Perrone2022}. The largest scale excited by MTI turbulence (the peak of the kinetic energy spectrum, or the integral scale) is set by a balance between energy injection by the MTI and the stabilizing entropy stratification, hence the moniker of buoyancy scale, and on theoretical grounds was predicted to scale as\footnote{However, in the three-dimensional simulations of \citet{Perrone2022a} the scaling of $l_\rmn{i}$ with $\chi$ was found to be somewhat shallower than $1/2$, a discrepancy which was attributed to the limited scale separation between injection and viscous dissipation.}
\begin{align}
        l_\rmn{i} \propto \dfrac{(\chi \omega_{\rmn{T}})^{1/2}}{N}.
\end{align}
A second important scale in MTI turbulence is the conduction length 
\begin{align}
        l_\chi \equiv (\chi / \omega_{\rmn{T}})^{1/2}, \label{eq:conduction_length}
\end{align} 
which represents how far heat can diffuse on timescales of $\omega_{\rmn{T}}^{-1}$, the turnover time of MTI turbulence. The integral and conduction scales differ by a factor of $ \omega_{\rmn{T}} / N$, which however is of order unity in the periphery of galaxy clusters, and the two are thus roughly comparable. Turbulence on scales below the integral scale is characterized by vertically elongated eddies excited by the buoyancy force. These eddies are efficient at transporting heat through advection and they drive an SSD to approximate equipartition with the kinetic energy. On scales $\gtrsim l_\rmn{i}$, the stabilizing effect of the entropy stratification is stronger, and the turbulent motions become relatively isotropic. On even longer scales, vertical motions are effectively suppressed by the stable entropy stratification.
Finally, combining the scalings of the rms velocities with the integral scale allowed \citetalias{Perrone2022a} to define a Reynolds number of MTI turbulence $\text{Re}_\rmn{i}$ as
\begin{align}
        \text{Re}_\rmn{i} \propto \dfrac{\chi \omega_{\rmn{T}}^2}{\nu N^2}.
\end{align}
For typical ICM conditions in the periphery of galaxy clusters and assuming unsuppressed to moderately suppressed thermal diffusivity with respect to the Spitzer value (Eq.~\ref{eq:spitzer-diffusivity}), the MTI is capable of driving turbulent motions of several hundred kilometers per second, on scales of $\sim 100 \, \si{kpc}$, thus potentially contributing to the observed levels of turbulence \citepalias{Perrone2022a}.
On the other hand, if the thermal diffusivity is significantly reduced by whistler wave scattering, the linear excitation and nonlinear saturation of the MTI could be strongly impacted. In particular, \citet{Drake2021} argued that due to the loss of isothermality along magnetic field lines caused by whistler suppression, the range of scales potentially MTI-unstable could be severely restricted to well below $\sim 20 \, \si{kpc}$. We believe these assessments to be rather on the conservative side, since it is likely that the ICM is in a transitional regime between the collisional Spitzer heat flux and the marginal whistler flux (see Sect.~\ref{sec:reference_parameters}), and the MTI can grow at a substantial fraction of its maximum rate even when conduction and buoyancy timescales are comparable. Nevertheless,
whistler suppression can slow down or even prevent the growth of an MTI eigenmode, a scenario that we study more in detail in Appendix~\ref{app:linear_theory}. In the next sections, instead, we  mostly deal  with the nonlinear saturation of the MTI with whistler suppression.

\subsection{Governing equations}

In this work we adopt the Boussinesq approximation to simulate the subsonic and small-scale dynamics of fluctuations within a quasi-spherical cluster characterized by a background density $\rho_0$, pressure $p_0$, and temperature $T_0$. By small scales, we here mean on scales much smaller than the typical pressure (or density) scale-height in the periphery of galaxy clusters, typically several hundred $\si{kpc}$. The main advantage of the Boussinesq approximation is that the fast-propagating sound waves are screened out of the fluid equations, leading to a considerable computational speed-up. As a result, the fluid is in pressure equilibrium at all times, and temperature and density fluctuations are related by
\begin{align}
        \frac{\delta T}{T_0} = - \frac{\delta \rho}{\rho_0}.
\end{align}
Our model describes a small Cartesian block of ICM plasma located at a given spherical radius $R_0$, where the vertical direction is along the radius and antiparallel to the gravitational acceleration $\bm g = - g_0 \bm e_z$. 
The background radial stratification in both entropy and temperature is included via the (constant-in-$z$) squared Brunt-V\"ais\"al\"a, $N^2$, and squared MTI,  $\omega_{\rmn{T}}^2$, frequencies. Their values are computed at the reference radius, $R_0$, using Eq.~\eqref{eq:freq}.
For further discussion on the specifics and limitations of the Boussinesq approximation, see \citetalias{Perrone2022}.

The Boussinesq variables evolved in our simulations are the velocity fluctuation $\bm u$, the magnetic field $\bm B$ (rescaled by $\sqrt{4 \pi \rho_0}$ so that it has the dimensions of a velocity), and the buoyancy variable $\theta$,
\begin{align}
   \theta  \equiv g_0 \frac{\delta \rho}{\rho_0}, \label{eq:theta_def}
\end{align}
(in units of $\si{cm.s^{-2}}$). The appropriate equations are derived in \citetalias{Perrone2022}, and are shown hereafter including a spatially varying anisotropic thermal diffusivity and a driving random external acceleration $\bm f$ in the velocity equation. They read
\begin{align}
        \bnabla \bcdot \bm u = \bnabla \bcdot \bm B &= 0 ,\label{eq:div_eq} \\
        \left( \partial_t + \bm u \bcdot \bnabla \right) \bm u &= - \frac{\bnabla p_\text{tot}}{\rho_0}  - \theta \bm e_z + \left( \bm B \bcdot \bnabla \right) \bm B + \nu \bnabla^2 \bm u + \bm f, \label{eq:mom_eq} \\
        \left( \partial_t + \bm u \bcdot \bnabla \right) \bm B &= \left( \bm B \bcdot \bnabla \right) \bm u + \eta \bnabla^2 \bm B, \label{eq:bfield_eq} \\
        \left( \partial_t + \bm u \bcdot \bnabla \right) \theta &= N^2 u_z +  \bnabla \bcdot \left[ \chi \bm b \left( \bm b \bcdot \bnabla \right)\theta \right] +  \omega_{\rmn{T}}^2 \bnabla \bcdot \left( \chi \bm b b_z\right) \nonumber \\ 
    &+ \chi_{\text{hyp}} \bnabla^{6} \theta ,  \label{eq:buoyancy_eq}
\end{align}
where $p_\text{tot}$ is the sum of thermal and magnetic pressure, $\nu$ and $\eta$ are the fluid viscosity and magnetic resistivity (both in units of $\si{cm^2.s^{-1}}$), while in Eq.~\eqref{eq:buoyancy_eq} the thermal diffusivity $\chi$ (also in $\si{cm^2.s^{-1}}$) absorbs a factor of $(\gamma -1)/\gamma$. We note that in the buoyancy equation we have also included a sixth-order diffusion operator where $\chi_{\text{hyp}}$ is the corresponding hyperdiffusivity (in units of $\si{cm^6.s^{-1}}$). We refer to Appendix \ref{app:implementation_whistler_suppression} for a detailed description of our implementation of whistler suppression and a discussion of why hyperdiffusivity is required to regularize the smallest scales when whistler suppression causes strong spatial variations in the thermal diffusivity.

\subsection{Boussinesq model of whistler suppression}\label{sec:estimates_whistler_mti}

In this section we introduce the final form of the whistler-suppressed thermal diffusivity that we employ in our numerical model. In terms of Boussinesq variables, Eq.~\eqref{eq:whistler_suppression_approx} can be rewritten as  (see Appendix~\ref{app:chi_derivation} for further discussion)
\begin{align}\label{eq:whistler-diffusivity}
        \chi \equiv \dfrac{\chi_0}{1 + \frac{1}{3}  \alpha \tilde{\beta} \sigma },
\end{align}
where we introduced the following three nondimensional quantities:
\begin{align}
        &\alpha \equiv \frac{1}{\gamma} \left( \frac{\lambda_{\text{mfp,e}}}{H}\right) \left( \frac{H^2}{\chi_0  / \omega_{\rmn{T}}}\right) , \label{eq:alpha_def} \\ 
    	&\tilde{\beta} \equiv \left( \frac{2 \varv_{\mathrm{cond}}^2}{\varv_\rmn{A}^2 }\right), \label{eq:beta_tilde_def} \\
        &\sigma \equiv \left\lvert b_z + \frac{ \bm b \bcdot \bnabla \theta}{\omega_{\rmn{T}}^2} \right\rvert. \label{eq:isolength_def} 
\end{align}
In Eq.~\eqref{eq:alpha_def} the $\alpha$ parameter encodes the physical information of the volume of ICM plasma that we simulate. More precisely, it represents the combined effect of plasma collisionality (given by the ratio $\lambda_{\text{mfp,e}}/H$, the Knudsen number, where $H$ is the pressure scale-height) and of the efficiency of thermal conduction on large (cluster-size) scales over an MTI $e$-folding time, as expressed by the squared ratio of $H$ to the unsuppressed conduction length (see Eq.~\ref{eq:conduction_length}). This parameter is small when the plasma has large collisionality or is ineffective at conducting heat. Conversely, it is large when the plasma becomes less and less collisional, or when heat transport is efficient on large scales.  In our subgrid model for whistler suppression, $\alpha$ is the only free parameter and is fixed at the beginning of each run. The second quantity in Eq.~\eqref{eq:beta_tilde_def} is a modified plasma $\tilde{\beta}$ that plays the same role as the usual plasma beta, but with a conduction speed $\varv_{\mathrm{cond}}$, defined as 
\begin{align}
    \varv_{\mathrm{cond}} \equiv (\chi_0 \omega_{\rmn{T}})^{1/2} , \label{eq:conduction_speed}
\end{align}
which replaces the ion thermal velocity. As we show below, the two betas are roughly comparable.
Finally, in Eq.~\eqref{eq:isolength_def} the isothermality parameter $\sigma$ 
is a dimensionless quantity that represents the ratio of the background to the total (background $+$ perturbation) temperature gradient  parallel to the magnetic field. 
It can be understood as follows: assuming for simplicity purely radial magnetic fields, $\sigma$ is equal to unity at quiescence (no temperature fluctuations), while in a turbulent medium it is greater (smaller) than unity if the gradient of the local temperature fluctuation is aligned (anti-aligned) with the background gradient. When the two gradients exactly cancel each other ($\sigma = 0$), the plasma is locally isothermal. In MTI turbulence at saturation $\sigma$ is smaller than unity, since the MTI effectively enforces isothermality along magnetic field lines  \citepalias[particularly so on scales below the conduction length $l_{\chi}$, see][]{Perrone2022a}.
Much like $\tilde{\beta}$ is the Boussinesq counterpart of the usual plasma $\beta$, the product $\alpha \sigma$ plays the role of the ratio $\lambda_{\text{mfp,e}}/L_{T,\parallel}$ in Eq.~\eqref{eq:whistler_suppression_approx}.

\subsection{Reference parameters in the ICM}\label{sec:reference_parameters}

We now estimate reference values for the suppression parameter $\alpha$, and proceed to quantify the impact of whistler suppression of thermal diffusivity on the MTI and its saturated turbulent state.  
These estimates inform the choice of parameters of our simulations (Sect.~\ref{sec:initial-conditions}).
Using the collisional mean-free path (Eq.~\ref{eq:lambda_mfp_spitzer}) and the Spitzer diffusivity (Eq.~\ref{eq:spitzer-diffusivity}) we estimate the suppression parameter $\alpha$ to be on the order of
\begin{align}
    \alpha \simeq 0.01 \left( \frac{H}{300\,\si{kpc}} \right) \left( \frac{\omega_{\rmn{T}}}{(700 \, \si{Myr})^{-1}} \right) \left( \frac{T_\rmn{e}}{5 \, \si{keV}} \right)^{-1/2} \label{eq:alpha_numbers} .
\end{align}
To evaluate $H$ we use the best-fit profiles for the X-COP sample reported by \citet{Ghirardini2019}, and take a fiducial radius (where $H$ is to be computed) on the order of several hundred $\si{kpc}$. By doing so, we find that the realistic range of $\alpha$ allowed by observations is 0.01--0.05, where the lower estimate is obtained from Eq.~\eqref{eq:alpha_numbers} with the values given, while the higher estimate is calculated assuming $H = 600\,\si{kpc}$, $\omega_\rmn{T} = (600 \, \si{Myr})^{-1}$ and $T_\rmn{e} = 1 \, \si{keV}$. It is important to note that $\alpha$ can have a large spread in values between different regions of the cluster periphery, and across different clusters. Nevertheless, we believe that the quoted numbers represent a realistic order-of-magnitude estimate of $\alpha$ in the ICM.

The modified $\tilde{\beta}$ can be similarly estimated using Eq.~\eqref{eq:conduction_speed} and the Spitzer diffusivity (Eq.~\ref{eq:spitzer-diffusivity}),
\begin{align}
    \tilde{\beta} \simeq 24 \left( \frac{\omega_{\rmn{T}}}{(700 \, \si{Myr})^{-1}} \right) \left( \frac{T_\rmn{e}}{5 \, \si{keV}} \right)^{5/2} \left(\frac{B}{2 \,\mu\si{G.}} \right)^{-2},
\end{align}
which could vary by up to an order of magnitude ($\tilde{\beta} \simeq 2-24$, where the lower bound was computed assuming $T_\rmn{e} = 1\,\si{keV}, B = 1\,\si{\mu G}$) due to the strong dependence on the temperature and on the magnetic field strength. Assuming typical conditions in the periphery of galaxy clusters ($n_\rmn{e} = 10^{-3} \si{cm^{-3}}, T_\rmn{e}= 5\,\si{keV}, B = 2\,\si{\mu G}$), $\tilde{\beta}$ is approximately $1/2$ of the electron plasma $\beta$ (see Table~\ref{tab:parameters}). 
Since for MTI turbulence $\sigma$ is less than unity, we can expect that the overall suppression of thermal diffusivity by whistler waves is relatively modest, with an average diffusivity on the order of $71-99 \%$ of the full Spitzer value (assuming conservatively $\sigma = 1$), which is in agreement with the estimates presented in \citet{Komarov2018}. While this level of suppression may not appear particularly significant, we note that these estimates do not take into account the nonlinear backreaction of whistler suppression onto the MTI turbulence: as we show in the next sections, even mild levels of suppression can have a strong impact on MTI-driven turbulence. A summary of all the definitions and the numerical values of the quantities introduced in this Section can be found in Table~\ref{tab:parameters}.

\begin{table}  
	\centering 
    \caption{Physical parameters of the ICM in the periphery of galaxy clusters. 
    }
    \label{tab:parameters} 
    \begin{tabular}{lp{20mm}p{42mm}}
	    \hline \hline
	    Quantity & Definition  & Value \\
	    \hline \hline
	    $\chi_{0}$ & Eq.~\eqref{eq:spitzer-diffusivity} & $ 1.65 \times 10^2 \; \si{kpc^2.Myr^{-1}}$ \\
   		\hline
	    $\omega_{\rmn{T}}$ & Eq.~\eqref{eq:freq} & $(600 \, \si{Myr})^{-1} \simeq 5.3\times 10^{-17} \si{s^{-1}}$ \\
	    $N$ & Eq.~\eqref{eq:freq} & $\simeq \omega_{\rmn{T}}$ \\
	    \hline
   		$\rho_{\text{e}}$ & Eq.~\eqref{eq:gyroradius_electrons} & $2.73\times 10^{-14}\, \si{kpc}$ \\
   		$\lambda_{\text{mfp,e}}$ & Eq.~\eqref{eq:lambda_mfp_spitzer} & $5.83\, \si{kpc}$ \\
   		$l_{\chi}$      & Eq.~\eqref{eq:conduction_length}$^\ddag$  & $320\, \si{kpc}$ \\
   		$H$ & $\lvert d \ln p / d R \rvert^{-1}$ & $300-600\, \si{kpc}$\\
   		\hline
	    $c_\rmn{s}$ & $ (\gamma T / \mu m_\rmn{p})^{1/2}$ & $890 \, \mu^{-1/2} \, \si{km.s^{-1}}$ \\
	    $\varv_{\mathrm{cond}}$ & Eq.~\eqref{eq:conduction_speed}$^\ddag$ &  $510 \, \si{km.s^{-1}}$\\
	    $\varv_\rmn{A}$ & $B/(4\pi \rho)^{1/2}$ & $140 \, \si{km.s^{-1}}$ \\
   		\hline
	    $\beta_\rmn{e}$ & $8 \pi n_\rmn{e}  T_\rmn{e}/B^2$  & $50$ \\
	    $\mathrm{Kn}$   & $\lambda_{\text{mfp,e}} / H $ & $0.01-0.02$ \\
   		\hline
	    $\alpha$ & Eq.~\eqref{eq:alpha_def} & $0.01-0.05$ \\
   		$\tilde{\beta}$ & Eq.~\eqref{eq:beta_tilde_def} & $2-24$ \\
	    $\sigma$ & Eq.~\eqref{eq:isolength_def} & $\lesssim 1$ (in MTI turbulence) \\
	    \hline \hline
    \end{tabular}
    \tablefoot{
    	To compute the estimates above, we assumed quasi-neutrality ($n_\rmn{e} \simeq n_\rmn{i}$) and equal temperature between ions and electrons ($T_\rmn{e} \simeq T_\rmn{i} \simeq T$). Moreover, we took as reference parameters $n_\rmn{e} = 10^{-3} \si{cm^{-3}}, T_\rmn{e} = 5\,\si{keV}, B = 2\,\si{\mu G}$. For the suppression parameter $\alpha$ and the modified plasma beta $\tilde{\beta}$, instead, we quote the realistic range of values that they could take in the periphery of galaxy clusters, as discussed in the main text. The values of $\omega_{\rmn{T}}$ and $N$ were taken from \citetalias{Perrone2022}, using data from \citet{Ghirardini2019}. The $^\ddag$ symbol means that the quantity has been evaluated assuming diffusivity equal to the Spitzer value. To compute $H$ we assumed a power-law profile with slope $-2$ to $-3$ and evaluated it at $R = 1\,\si{Mpc}$.  
    }
\end{table}

\section{Methods}\label{sec:numerical_methods}
In this section we describe the numerical methods that we have used, alongside the diagnostics, the initial conditions and the parameters of our simulations.

\subsection{\textsc{SNOOPY} code}
\label{snoopy}

The simulations presented in this work have been performed with a modified version of SNOOPY, a pseudo-spectral three-dimensional (3D) MHD code \citep{Lesur2015}, where we included anisotropic heat conduction and implemented our closure for the whistler-suppressed heat diffusivity (Eq.~\ref{eq:whistler-diffusivity}). In the SNOOPY code, a Fourier transform is applied to the MHD equations, and the Fourier coefficients of each wavenumber $\bm k = k_x \bm{e}_x + k_y \bm{e}_y + k_z \bm{e}_z$ are then advanced in time with a three-step Runge-Kutta algorithm. In addition to employing a standard $2/3$ de-aliasing rule to all the physical variables, we decide to filter the whistler-suppressed diffusivity through a two-point harmonic averaging and to apply a hyperdiffusion operator to the buoyancy equation (Eq.~\ref{eq:buoyancy_eq}) to avoid Gibbs oscillations at the grid-scale. For further details on the numerical implementation, see Appendix~\ref{app:implementation_whistler_suppression}.

\subsection{Diagnostics}

We use a variety of diagnostics to explore the impact of whistler suppression on the MTI, both in real and in spectral space. We track the time evolution of the kinetic ($K$), magnetic ($M$), and potential ($U$) specific energies, as follows:
\begin{align}
    K = \frac{1}{2} \langle u^2 \rangle, \hspace{1em} M = \frac{1}{2} \langle B^2 \rangle, \hspace{1em} U = \frac{1}{2 N^2} \langle \theta^2 \rangle.
\end{align}
Here the angle brackets denote the volume average (we remind the reader that here and in the next sections the magnetic field is rescaled by $\sqrt{4 \pi \rho_0}$ so that it has units of a velocity). We similarly track the mean suppression of thermal diffusivity, $\langle \chi / \chi_0 \rangle$. In addition to these volume-averaged quantities, we also follow the time-evolution of the energy contained at each Fourier wavenumber $k= \lvert \bm k \rvert$ (the so-called shell-integrated power spectra), as well as the rates of energy injection and removal at each given $k$ by all the various physical processes (e.g., fluid advection, magnetic tension, dissipative processes, and so forth) corresponding to the terms in Eqs.~\eqref{eq:mom_eq}-\eqref{eq:buoyancy_eq}. For instance, the evolution of the potential energy contained at each wavenumber is described by 
\begin{align}
    \frac{d }{d t} \left( \frac{1}{2 N^2} \lvert \hat{\theta} (\bm k, t) \rvert^2 \right) = \frac{1}{N^2} \sum_i \Re \left\{ \hat{\theta}^* (\bm k, t) \mathcal{F} \left[ w_i \right] \right\},
\end{align}
where $\mathcal{F}$ is the Fourier transform operator (also denoted by a hat $\hat{~}$), the asterisk denotes the complex conjugate, and $w_i$ are the terms corresponding to the fluid advection ($-\bm u \bcdot \bnabla  \theta$), the buoyancy force ($N^2 u_z$), the anisotropic dissipation ($\bnabla \bcdot \left[ \chi \bm b \left( \bm b \bcdot \bnabla \right)\theta \right]$), the energy injection rate from the background temperature gradient due to the MTI ($\omega_{\rmn{T}}^2 \bnabla \bcdot \left( \chi \bm b b_z\right)$), and the hyperdissipation ($ \chi_{\text{hyp}} \bnabla^{6} \theta$). The shell-averaged potential spectral energy density $E_U(k)$ is then computed as
\begin{equation}
        E_U(k,t) dk = \sum_{k \leq \lvert \bm k' \rvert < k + dk} \frac{1}{2 N^2} \lvert \hat{\theta} (\bm k', t) \rvert^2 ,
\end{equation}
and similarly for the kinetic and magnetic spectral energy densities.

\subsection{Initial conditions and parameters}\label{sec:initial-conditions}

\subsubsection{2D simulations}

We perform four two-dimensional (2D) runs with increasingly stronger whistler suppression ($\alpha = 4 \times 10^{-5}, 4 \times 10^{-4}, 2 \times 10^{-3}, 4 \times 10^{-3}$) and compare them to a reference unsuppressed ($\alpha = 0$) MTI run. The reason why we first focus on 2D runs is that despite the different saturation mechanism (which in 2D involves an inverse energy cascade to large scales) some key features of MTI turbulence are similar between 2D and 3D \citepalias{Perrone2022,Perrone2022a}. One notable exception is the absence of an SSD in 2D \citep{Zeldovich1957}, which is shown to play a crucial role in the whistler suppression of MTI turbulence. For this reason we believe that 2D simulations constitute a useful benchmark which helps us investigate and isolate the main features of whistler suppression and the impact of the SSD.

It should be noted that in the 2D runs, the values of the suppression parameter $\alpha$ are lower than what was estimated in Sect.~\ref{sec:reference_parameters}. This is to obviate the fact that in 2D without an SSD, the magnetic energy at saturation (which affects the level of whistler suppression according to Eq.~\ref{eq:whistler-diffusivity}) is not pinned at equipartition with the kinetic energy, but depends on the initial strength of the imposed (weak) uniform magnetic field. Boosting the initial uniform $\bm B$ field to match the equipartition strength is also problematic, since in 2D strong fields lead to formation of structures at the box size \citepalias{Perrone2022a}. To get around this issue, for our 2D simulations only we decide to compensate the unphysically larger value of $\tilde{\beta}$ with a correspondingly lower $\alpha$, such that their product (and therefore the overall suppression of thermal conduction, see Eq.~\ref{eq:whistler-diffusivity}) roughly coincides with the levels estimated in Sect.~\ref{sec:reference_parameters}. We believe that this strategy allows us to obtain a more realistic picture of whistler suppression on the MTI, and we note that this issue is fully overcome with our 3D simulations of Sect.~\ref{sec:three-dimensional-sims}.

All runs have a resolution of $1024^2$ and aspect ratio unity, thermal diffusivity $\chi_0 / (L^2 \omega_{\rmn{T}}) = 10^{-3}$ with a thermal Prandtl number $\text{Pr} = \nu/\chi =  0.01$, and a magnetic Prandtl number $\text{Pm} =  \nu/\eta =  1$. The hyperdiffusivity is $\chi_{\mathrm{hyp}} / (\Delta h^4 \chi_0) = 5.2 \times 10^{-3}$, where $\Delta h = L / 1024$ the grid size. The runs are initialized with a mean uniform horizontal magnetic field with strength $B_0 / (L \omega_{\rmn{T}}) = 10^{-4}$, and random velocity fluctuations of amplitude $|\delta u_i | / (L \omega_{\rmn{T}}) = 10^{-4} $, for $i = x,z$. The magnetic field lines are initially isothermal ($\sigma = 0$) and the thermal diffusivity does therefore not feel any whistler suppression at time $t=0$.

\begin{figure*}
        \centering
        \includegraphics[width=1.0\linewidth]{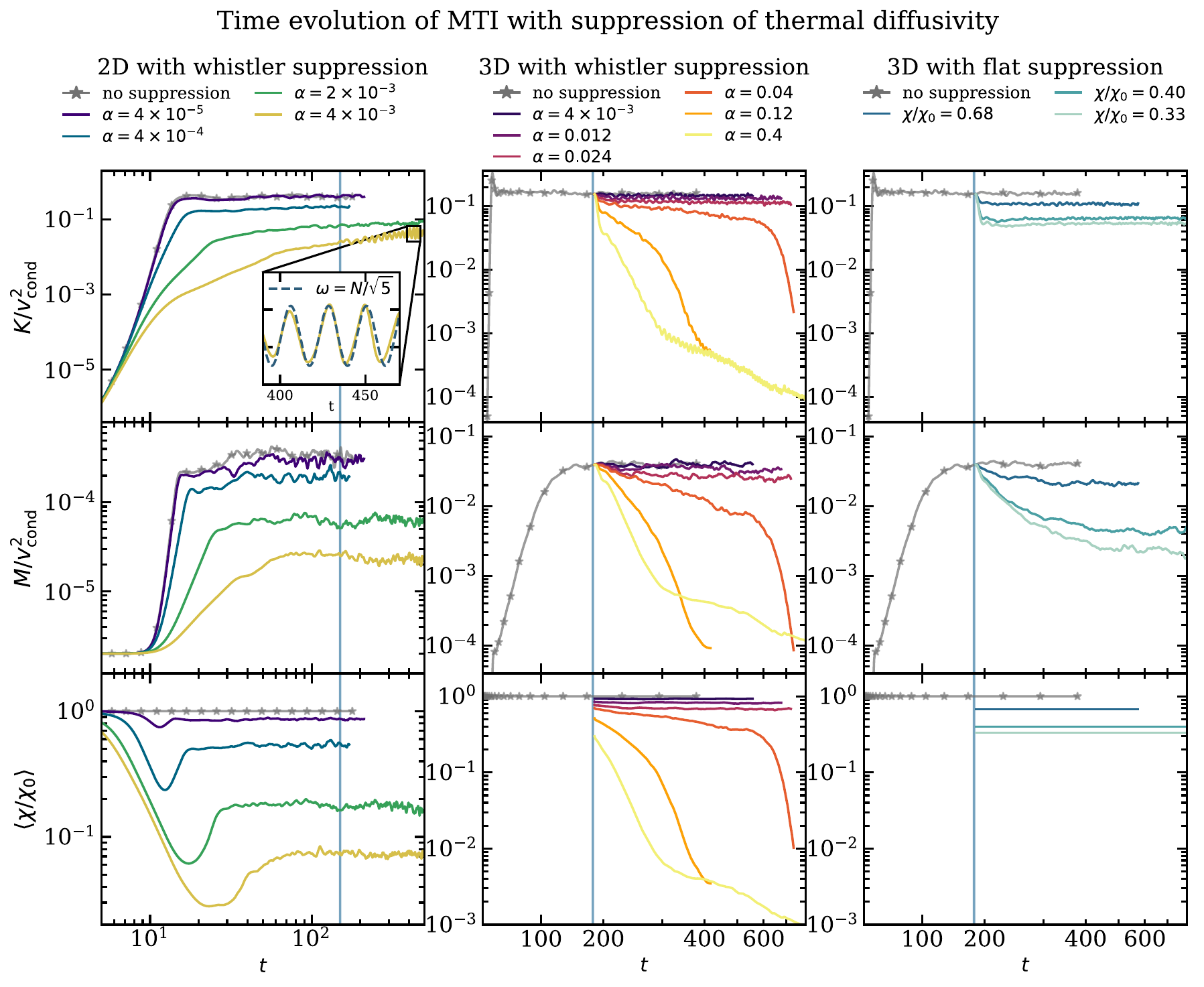}
        \caption{Time evolution of the MTI runs with whistler suppression of thermal diffusivity in 2D and in 3D (left and middle columns, respectively), and in 3D with flat suppression (right column). From top to bottom: Volume-averaged kinetic energy, magnetic energy, and average suppression factor as a function of time. In 3D with whistler suppression, MTI turbulence decays if the suppression parameter $\alpha$ is large enough, while in 2D (where there can be no SSD) or in 3D with flat uniform suppression (with or without an SSD) MTI turbulence never dies completely.
    	Left panel: The cyan vertical line at $t=150$ indicates the time of the snapshots shown in Fig.~\ref{fig:mti2D_final_states}. The inset in the top panel zooms in on the simulation with $\alpha = 4 \times 10^{-3}$ (gold solid line) at late times, and shows periodic oscillations of frequency $\omega = N/\sqrt{5}$ associated with a dominant large-scale $g$ mode $(k_x,k_z) = (1,2) \times 2\pi/L$. Middle and right panels: The cyan vertical line at $t=186$ indicates the time when we turn on suppression of thermal diffusivity in our reference run. 
    	}
        \label{fig:time_evolution_mti_2d_3d}
\end{figure*}

\subsubsection{3D simulations}\label{sec:three-dimensional-sims}

In addition to our 2D runs, we perform a number of 3D simulations with whistler suppression. This allows us to examine a more realistic regime where the magnetic field strength is kept self-consistently at equipartition by the SSD, and feeds back into the suppression of thermal conductivity.

We perform six 3D runs with whistler suppression and increasing $\alpha$ between $[4 \times 10^{-3}, 0.4]$, and compare the results to a reference 3D run without suppression ($\alpha = 0$). The values of $\alpha$ are chosen so as to sample the range that can be realistically expected in the periphery of galaxy clusters (see Sect.~\ref{sec:reference_parameters}). All our 3D runs have a resolution of $288^3$, aspect ratio of unity, and $\chi_0 / (L^2 \omega_{\rmn{T}}) = 0.0025$. To reflect the hierarchy between thermal, viscous and resistive diffusivities in the ICM, we choose explicit viscosity and resistivity such that $\text{Pr} \simeq 0.042$ and $\text{Pm} = 4$. The hyperdiffusivity is $\chi_{\mathrm{hyp}} / (\Delta h^4 \chi_0) = 4.8 \times 10^{-3}$. 

In order to save computational time, we simulate the linear phase of the instability and the subsequent kinematic and nonlinear dynamo phase only once in our reference run. Then, once it reaches saturation, we choose one snapshot and use it as a starting point for our series of runs with whistler suppression. The reference run is initialized with random velocity fluctuations of amplitude $|\delta u | / (L \omega_{\rmn{T}}) = 10^{-4} $ in all directions. Contrary to our 2D runs, we do not include a mean uniform horizontal magnetic field, but start with a helicoidal field varying with $z$ pointing in the $x-y$ plane: $\bm B = B_0 \left[  \sin(2 \pi z / L)\bm e_x + \cos(2 \pi z / L)\bm e_y \right]$, with $B_0 / (L \omega_{\rmn{T}}) = 10^{-4}$.

We stress that the lack of a net magnetic flux in our 3D simulations means that -- if not replenished by an SSD -- magnetic fields are now allowed to resistively decay to negligible values. This introduces a new equilibrium state available to the system where magnetic fields are zero, thermal conduction completely suppressed and the MTI rendered ineffective (a ``dead'' state). Assessing whether and under what conditions the system reaches this dead state is one of the main topics of the following sections. We decide not to run 3D simulations with a (weak) net magnetic flux because, if they are able to sustain an SSD, we expect that their final state at saturation will show little difference from an equivalent no-net flux run (as shown in \citetalias{Perrone2022a}); if they cannot sustain an SSD, on the other hand, then the residual thermal diffusivity after the turbulent magnetic fields decay will depend on the initial arbitrary weak flux, and will not be representative of the physical conditions in the periphery of galaxy clusters.

\subsection{Simulations with external forcing}\label{sec:numerics-external-forcing}

To study the impact of additional sources of turbulence to the MTI with whistler suppression, we run a number of MTI simulations where we add an external forcing to the momentum equation. In this paper we show the results of two MTI runs with external forcing, one without whistler suppression ($\alpha=0$), and another with strong whistler suppression ($\alpha = 0.12$). For comparison, we also run a forcing-only run (without the MTI) with isotropic thermal conductivity equal to $\chi_0$ and no whistler suppression, while keeping all other parameters unchanged.

These simulations are the same as in \citetalias{Perrone2023}, and are initialized from a snapshot at late times of the corresponding 3D run without external forcing introduced in Sect.~\ref{sec:three-dimensional-sims}. We report here the numerical details for completeness.
We implement a delta-correlated in time, solenoidal,  white-noise forcing in Fourier space by drawing six random numbers at each timestep for every wavevector $\bm k$ in a specified range (we choose $|\bm k| =  \{ 4,5,6 \} \times 2 \pi / L$): three random numbers determine an auxiliary random vector used to enforce the divergence-free condition ($\bm k \bcdot \hat{\bm f}_{\bm{k}} = 0$), and three multiply the complex phases of the forcing vector $\hat{\bm f}_{\bm{k}}$. The amplitude of $\hat{\bm f}_{\bm{k}}$, instead, is constant and the same for all the forced wavevectors. The resulting forcing has zero-mean helicity but allows for small fluctuations around zero. 

\subsection{Runs with flat uniform suppression}

In addition to whistler-suppressed thermal diffusivity, we also run a number of simulations with a flat uniform suppression below the Spitzer value
\begin{align}
    \chi = f \chi_0,
\end{align}
where $f < 1$ is a constant. Comparing the two models allows us to clearly identify the features of whistler suppression and what physics might be missed by employing a simple flat uniform suppression.

\section{Weakening of the MTI and critical suppression}\label{sec:weakening-suppression}

In this section, we show that both in 2D and in 3D simulations the MTI turbulence is weakened as the suppression parameter $\alpha$ increases (Sect.~\ref{sec:time_evolution_2d_3d}). However, while in 2D we observe a gradual decrease in the turbulent levels as we sweep through the range of $\alpha$, in 3D there is a sharp transition at a critical value of the suppression parameter, beyond which the MTI turbulence dies (Sect.~\ref{sec:2d-3d-suppression}). We relate this behavior to the impairment of the SSD, and we propose a simple model to predict the critical $\alpha$ (Sect.~\ref{sec:ssd_shutdown}).

\subsection{Time evolution of the MTI with whistler suppression}\label{sec:time_evolution_2d_3d}

In Fig.~\ref{fig:time_evolution_mti_2d_3d} we show the time-series of the volume-averaged kinetic, potential and magnetic energy and the average suppression factor of our whistler-suppressed runs in 2D (left column), 3D with whistler suppression (middle column), and 3D with uniform suppression (right column). 

In 2D (where we evolve the MTI with whistler suppression throughout the linear phase of the instability), we observe that simulations with larger $\alpha$ show a shorter phase of exponential growth, followed by a secondary growth phase until the system finally saturates. We also note that the average suppression factor decreases sharply during the linear phase of the instability before eventually stabilizing at somewhat larger values. This is due to the fact that as the instability grows the magnetic field line is brought further away from isothermality ($\sigma$ increases). However, as the MTI approaches the nonlinear phase and the vertical magnetic perturbation grows stronger, the total magnetic energy increases (see second row of Fig.~\ref{fig:time_evolution_mti_2d_3d}, left column), $\tilde{\beta}$ decreases and the whistler suppression becomes less effective. For further details concerning the nonlinear saturation of the MTI with whistler suppression, see Appendix~\ref{app:linear_theory}.

After the instability has saturated, we observe that in both 2D and 3D with whistler suppression and with the lowest $\alpha$ there is little difference compared to the reference MTI run, and the final energies are similar.
With higher suppression parameter the behavior between 2D and 3D drastically differs: while in 2D the final saturated energies progressively decrease as $\alpha$ increases, in 3D this gradual trend holds only for the three runs with the lowest suppression parameter ($\alpha = 4\times 10^{-3}, 0.012, 0.024$), beyond which the system undergoes a sharp transition and the turbulence dies out. This complete switch-off of turbulence is particularly evident in the run with $\alpha = 0.04$ (dark orange), which experiences a slow decline over hundreds of dynamical times before suddenly decaying. This behavior is visible across all the volume-averaged quantities tracked in Fig.~\ref{fig:time_evolution_mti_2d_3d} and further increasing $\alpha$ only accelerates the decay. This is in contrast to simulations with a model of flat uniform suppression of thermal conductivity (third column of Fig.~\ref{fig:time_evolution_mti_2d_3d}), where irrespective of whether the turbulence is sufficient to drive an SSD ($\chi / \chi_0 = 0.40, 0.68$) or not ($\chi / \chi_0 = 0.33$), the MTI turbulence never dies completely but is only reduced.

We interpret the radically different behavior with small changes to the $\alpha$ parameter as a sign of a critical transition in whistler-suppressed MTI turbulence. This is a feature of many nonlinear systems and we speculate that two dynamical fixed points exist for low values of $\alpha$, corresponding to the dead state and the MTI-turbulent state. As $\alpha$ is increased above a critical value the MTI turbulent state ceases to exist, and the system moves onto the dead state. A more in-depth analysis of this critical transition is carried out in Sect.~\ref{sec:ssd_shutdown}. This picture may change substantially in simulations with a (weak) net magnetic flux, which would set a lower limit for the thermal conductivity, thus removing the dead state. We speculate that, depending on the strength of the initial uniform magnetic field, the simulations may either become effectively dead (if the residual thermal diffusivity is too low for the MTI to grow), or possibly exhibit a cyclic behavior of MTI growth and decay.

Before proceeding with the analysis of the turbulence at saturation, we note that in the 2D simulations with largest $\alpha$ the system starts exhibiting an oscillatory pattern at late times, most clearly visible in the kinetic energy of Fig.~\ref{fig:time_evolution_mti_2d_3d}, left column. We explain this behavior with the emergence of a large-scale $g$ mode on top of the smaller-scale MTI turbulence, which we are able to track over a few cycles in the inset in Fig.~\ref{fig:time_evolution_mti_2d_3d} (top left panel). In this particular case the $g$ mode has wavevector $(k_x,k_z) = (1,2) \times 2\pi/L$, but in different runs we detected other orientations. The appearance of large-scale $g$ modes in MTI simulations is in itself not surprising, and indeed in 2D the instability saturates by exciting large-scale $g$ modes that block the inverse cascade of kinetic energy and redistribute it in the form of density fluctuations \citepalias{Perrone2022}. The difference here is that with strong whistler suppression $g$ modes end up dominating energetically over the smaller-scale turbulence. Interestingly, we observe the emergence of large-scales $g$ modes also in the decaying 3D runs, possibly excited by the weak residual MTI energy injection, meaning that the dead solution is not completely dead. However, we believe this phenomenon to be only transient given the ongoing decrease of the thermal diffusivity which further quenches MTI injection.

\begin{figure*}
        \centering
        \includegraphics[width=0.7\linewidth]{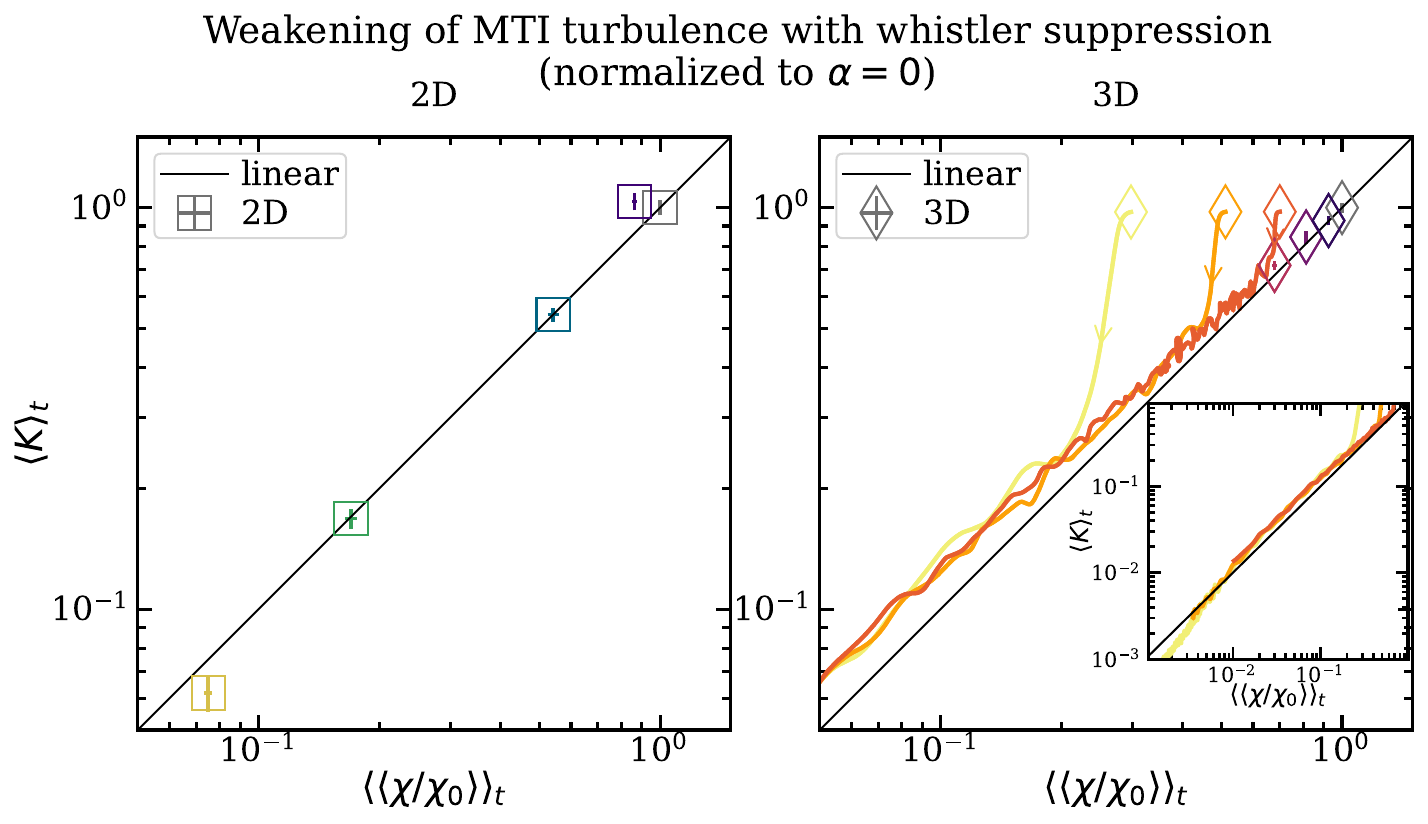}
        \caption{Time-averaged kinetic energy vs. mean suppression factor for our 2D simulations (left panel) and 3D simulations (right panel). The points are color-coded as in Fig.~\ref{fig:time_evolution_mti_2d_3d}. In 2D the time-averages were performed over a window of $t = [130, 170]$ to reduce contamination by the large-scale $g$ modes. In 3D we show time-averages for the three surviving runs only, while the three decaying runs are shown as solid lines (the full extent of the decay is shown in the inset). In both panels the black solid line represents a relation of linear proportionality. The linear proportionality between  conductivity and turbulence levels found in \citetalias{Perrone2022} and \citetalias{Perrone2022a} is seen to remain valid in an average sense in simulations including whistler-suppression of thermal conduction.}
        \label{fig:mti2D-3D_suppression_scalings}
\end{figure*}

\subsection{Correlation of kinetic energy with thermal diffusivity}\label{sec:2d-3d-suppression}

We now take a closer look at how the suppression of MTI turbulence relates to the volume-averaged $\chi / \chi_0$. In Fig.~\ref{fig:mti2D-3D_suppression_scalings} we plot the time-averaged kinetic energy versus the suppression factor for the 2D runs (left panel) and the 3D runs (right panel). We note that in 3D we compute the time-averages (denoted by $\langle ... \rangle_t$) only for the three surviving runs ($\alpha = 4\times 10^{-3}, 0.012, 0.024$) and we instead show the correlation of kinetic energy with thermal diffusivity as a function of time for the three runs that die ($\alpha = 0.04, 0.12, 0.4$). The different datapoints (each corresponding to a value of $\alpha$) closely follow a linear trend with the averaged $\chi / \chi_0$ both in 2D and in 3D. Interestingly, in 3D this trend applies not only to the runs that adjust to the new turbulent saturated state, but also to the runs that die as they decay. In other words, despite the strong spatial inhomogeneity of the effective thermal diffusivity $\chi/\chi_0$ (which is the subject of Sect.~\ref{sec:flow-backreaction}), in our simulations the turbulent kinetic energy appears to scale proportionally with the volume-averaged suppression factor, extending the basic scaling relationship of MTI turbulence between $K$ and $\chi$ in the whistler-suppressed regime. 
This finding lends weight to the idea that, at least as a first approximation, the impact of whistler suppression of thermal conductivity on the MTI can be described by a simple model with a flat uniform suppression. However, as we show in the next Section, a flat uniform suppression cannot explain the appearance of a critical transition in 3D MTI turbulence that we witnessed in Fig.~\ref{fig:time_evolution_mti_2d_3d}.

\begin{figure}[!ht]
        \centering
        \includegraphics[width=0.85\linewidth]{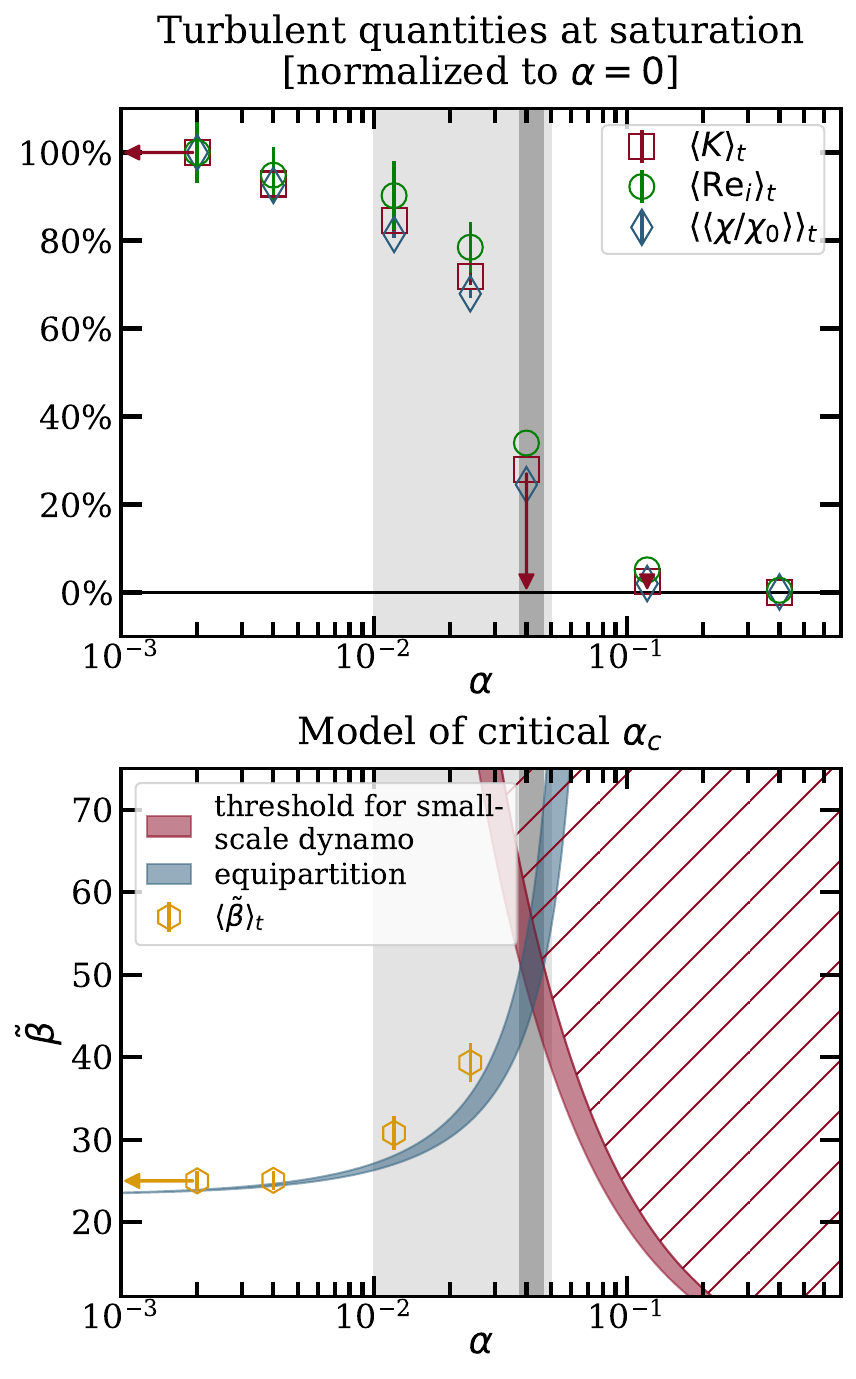}
        \caption{Critical transition to the dead state with whistler suppression. Top panel: Time-averaged turbulent kinetic energies, Reynolds number at the integral scale, and average suppression factor for our 3D runs as a function of $\alpha$. The values are normalized to those of the unsuppressed run. The leftmost point, corresponding to $\alpha=0$, was placed at $\alpha = 2\times 10^{-3}$ for visualization purposes. The three rightmost data points represent decaying runs, thus we do not show error bars. The  light gray shaded region indicates the approximate range of the $\alpha$ parameter for typical ICM parameters (see Table~\ref{tab:parameters}), while the dark gray shaded region represents the predicted critical $\alpha$ where the transition from MTI turbulence to dead state occurs. Bottom panel: Model for critical $\alpha$ with whistler suppression (see text). The blue band indicates the equipartition value of $\tilde{\beta}$ with the MTI turbulence (Eq.~\ref{eq:equipartition-crit}). The red band divides the $\alpha-\tilde{\beta}$ space into two regions: where an SSD is possible (below the curve) and where the SSD is not sustainable (hatched region) (see Eq.~\eqref{eq:ssd-criterion}). The two bands cross approximately at $\alpha \simeq 0.038-0.046$ above which MTI turbulence dies. The gold symbols show the saturated values of $\tilde{\beta}$ for the nondecaying runs.}
        \label{fig:mti3D_model_critical_alpha}
\end{figure}

\subsection{Impairment or shutdown of the SSD in 3D and the death-spiral}\label{sec:ssd_shutdown}

In this section we propose a simple toy model to explain the critical transition in 3D between weakened MTI turbulence with whistler suppression and the dead state. 
Our model rests on the assumption that at saturation MTI-driven turbulence pins kinetic and magnetic turbulent fluctuations approximately at equipartition 
\begin{align}\label{eq:equipartition-condition}
    \begin{tabular}{>{\raggedright\arraybackslash}p{0.45\linewidth}  >{\raggedright\arraybackslash}p{0.35\linewidth}}
      $\varv_{\rmn{A}}^2 \sim u_{\mathrm{rms}}^2$   & (equipartition), 
    \end{tabular}
\end{align}
provided that the criterion for an SSD \citepalias[magnetic Reynolds number $\mathrm{Rm} \equiv \mathrm{Pm} \times \mathrm{Re}_\rmn{i}$ larger than a critical value,][]{Perrone2022a} is fulfilled: 
\begin{align}\label{eq:ssd-condition}
    \begin{tabular}{>{\raggedright\arraybackslash}p{0.45\linewidth}  >{\raggedright\arraybackslash}p{0.35\linewidth}}
      $\mathrm{Rm}  \gtrsim \mathrm{Rm}_\rmn{c} \approx 35$   & (SSD).
    \end{tabular}
\end{align}
More precisely, our hypothesis is that the critical transition emerges when these two conditions cease to be satisfied: if magnetic fields cannot be kept at equipartition levels by an SSD, their decay leads to further suppression of thermal conductivity through its dependence on the plasma $\beta$. Stronger whistler suppression results in weaker MTI turbulence, and an even faster decay of magnetic energy, producing a self-reinforcing cycle which leads to the dead state and which we call the ``death-spiral''. 

The previous argument can be made more quantitative using the two conditions (Eqs.~\ref{eq:equipartition-condition}-\ref{eq:ssd-condition}) combined with the MTI scaling laws found in \citetalias{Perrone2022,Perrone2022a} and summarized in Sect.~\ref{sec:mti-theory}, allowing us to estimate the critical value of $\alpha$. In the following, we 
make the simplifying assumption that the effective volume-averaged suppression factor $\langle \chi/\chi_0 \rangle$ can be approximately expressed in terms of appropriate volume averages of $\tilde{\beta}$ and $\sigma$ as
\begin{align}\label{eq:fudge-factor}
        \left\langle \dfrac{\chi}{\chi_0} \right\rangle \approx \dfrac{1}{1 + \dfrac{\phi}{3} \alpha \langle \tilde{\beta} \rangle \langle \sigma \rangle },
\end{align}
where $\phi$  is a parameter that takes into account the nonlinear dependence of $\langle \chi/\chi_0 \rangle$ on $\tilde{\beta}$ and $\sigma$, their statistical correlation, and the shape of their distributions including intermittency (see Sect.~\ref{sec:flow-morphology}). We test Eq.~\eqref{eq:fudge-factor} comparing the predicted value of $\langle \chi/\chi_0 \rangle$ with the actual simulation data from the three surviving 3D runs, and find that the two sets of values match reasonably well for $\phi \simeq 3.0 - 3.7$ (the average relative error is $\sim 3 \%$).
With this caveat in mind, we drop the angle-bracket notation for the rest of the section and assume that we are always referring to volume-averaged quantities.

Using the MTI scaling law in Eq.~\eqref{eq:mti-kin-scaling}, the equipartition condition (Eq.~\ref{eq:equipartition-condition}) can be rewritten in terms of the specific magnetic energy as 
\begin{align}
    \frac{M}{\varv_{\text{cond}}^2} = \text{const.} \times \left\langle \dfrac{\chi}{\chi_0} \right\rangle \dfrac{\omega_{\rmn{T}}^2}{N^2},
\end{align}
where the constant of proportionality can be read directly from the value of $M/ \varv_{\text{cond}}^2$ at saturation for our reference 3D unsuppressed run in Fig.~\ref{fig:time_evolution_mti_2d_3d}, for which we have $M/ \varv_{\text{cond}}^2 \simeq 0.043$, and thus  $\text{const.} \simeq 0.043 N^2 / \omega_T^2 = 0.0043$. Replacing $\langle \chi/\chi_0 \rangle$ in the above equation with Eq.~\eqref{eq:fudge-factor}, and solving for $\tilde{\beta}$ 
the equipartition condition can be rewritten as
\begin{align}\label{eq:equipartition-crit}
    \begin{tabular}{>{\raggedright\arraybackslash}p{0.45\linewidth}  >{\raggedright\arraybackslash}p{0.35\linewidth}}
      $\dfrac{1}{\tilde{\beta}} = 0.0043\dfrac{\omega_{\rmn{T}}^2}{N^2} - \dfrac{\phi}{3} \alpha \sigma $   & (equipartition),
    \end{tabular}
\end{align}
where we used the equivalence $1/\tilde{\beta} \equiv M /\varv_{\mathrm{cond}}^2$.
The physical meaning of Eq.~\eqref{eq:equipartition-crit} is that, given a certain value of $\alpha$ (and of effective suppression of thermal diffusivity), the value of $\tilde{\beta}$ changes accordingly to maintain equipartition with the turbulent kinetic energy. Interestingly, the above equation already provides a constraint on the critical $\alpha$, as $1/\tilde{\beta}$ can never become negative.
By setting the l.h.s of Eq.~\eqref{eq:equipartition-crit} equal to 0 ($\tilde{\beta} \rightarrow \infty$), we can obtain an upper bound for $\alpha$:
\begin{align}\label{eq:alpha_inf}
    \alpha_{\rmn{max}} = \dfrac{0.0043 \omega_{\mathrm{T}}^2/ N^2}{\sigma \phi /3}.
\end{align}

A second independent constraint on $\alpha$ is provided by the criterion for the SSD which sets a lower limit on the magnetic Reynolds number below which the dynamo cannot be sustained. This constraint originates from the fact that in whistler-suppressed MTI turbulence the Reynolds number at the integral scale $\text{Re}_\rmn{i}$ also scales approximately linearly with the average suppression factor (Fig.~\ref{fig:mti3D_model_critical_alpha}, top panel)
\begin{align}\label{eq:reynolds-integral}
    \text{Re}_\rmn{i} \simeq  \text{Re}_{\rmn{i}0} \times \left\langle \dfrac{\chi}{\chi_0} \right\rangle,
\end{align}
where $\text{Re}_{\rmn{i}0}$ denotes the Reynolds numbers of the reference unsuppressed run.
Solving for $\tilde{\beta}$ and using Eq.~\eqref{eq:reynolds-integral}, the criterion for the SSD in Eq.~\eqref{eq:ssd-condition} can be recast as 
\begin{align}\label{eq:ssd-criterion}
    \begin{tabular}{>{\raggedright\arraybackslash}p{0.45\linewidth}  >{\raggedright\arraybackslash}p{0.35\linewidth}}
      $\tilde{\beta} < \dfrac{3}{\phi \alpha \sigma } \dfrac{ \mathrm{Rm}_{0} - \mathrm{Rm}_{\rmn{c}}}{\mathrm{Rm}_{\rmn{c}}}$   & (SSD),
    \end{tabular}
\end{align}
where $\mathrm{Rm}_{0} \equiv \mathrm{Pm} \mathrm{Re}_{i0}$ is the unsuppressed magnetic Reynolds number of the plasma.
The physical meaning of Eq.~\eqref{eq:ssd-criterion} is that in MTI turbulence with whistler suppression, only magnetic fields of strength above a certain $\alpha$-dependent threshold (equivalently, low enough plasma $\tilde{\beta}$) can be sustained self-consistently by the MTI-driven SSD. The two intersecting hyperbolae in the $\alpha-\tilde{\beta}$ plane define a critical value $\alpha_\mathrm{c}$. Because of the shape of the curves in Equations~\eqref{eq:equipartition-crit} and \eqref{eq:ssd-criterion}, the correct $\alpha_\mathrm{c}$ is given by the numerical solution to Equations~\eqref{eq:equipartition-crit} and \eqref{eq:ssd-criterion}, which is always bound to be less than the asymptotic value in Eq.~\eqref{eq:alpha_inf}.

If we plot the two conditions in Equations.~\eqref{eq:equipartition-crit} and \eqref{eq:ssd-criterion} (lower panel of Fig.~\ref{fig:mti3D_model_critical_alpha}, assuming a value of $\phi$ between $3.0-3.7$), we see that for very low values of $\alpha$, the equipartition $\tilde{\beta}$ (solid blue line) is well below the SSD threshold (solid red curve), and thus equipartition-level magnetic fields can be sustained by the SSD in presence of moderate suppression, as we observe in our 3D runs (golden hexagons in Fig.~\ref{fig:mti3D_model_critical_alpha}). As $\alpha$ is increased, however, the equipartition $\tilde{\beta}$ progressively nears the SSD threshold, and eventually crosses it: at this point, equipartition-level magnetic fields become too weak to sustain themselves, and the system enters the death-spiral. 
From our model, we estimate this critical parameter $\alpha_\rmn{c}$ to be on the order of $\alpha_\rmn{c} \simeq 0.038-0.046$ (depending on the value of the parameter $\phi$),\footnote{In drawing the lower panel of Fig.~\ref{fig:mti3D_model_critical_alpha} we have assumed a volume-averaged value of $\sigma=0.5$ which is constant with $\alpha$. This is in reasonable agreement with the results of our three surviving 3D runs (where the average $\sigma$ is $\left[ 0.52, 0.48, 0.46 \right]$ for increasing $\alpha$, respectively); using instead $\sigma=0.4$ the corresponding critical value is $\alpha_\rmn{c} \simeq 0.048-0.059$. }
which is consistent with the findings from our numerical simulations (where $\alpha_\rmn{c} \in \left( \left. 0.024,0.04 \right] \right.$).


In order to extrapolate to the physical conditions of the ICM, we note that the critical value of $\alpha$ for the impairment of the SSD resulting from Eq.~\eqref{eq:ssd-criterion} is a function of $\rmn{Rm}_0$: the higher the magnetic Reynolds number, the higher the allowed values of $\tilde{\beta}$ that can be self-sustained by a magnetic dynamo. Since in the ICM $\rmn{Rm}_0$ is likely very large (but undetermined), we can use Eq.~\eqref{eq:alpha_inf} to place an upper bound on $\alpha_\mathrm{c}$. 
Assuming $\omega_T^2 \simeq N^2$ (appropriate in the periphery of galaxy clusters) and $\sigma = 0.5$, $\phi \simeq 3.0 - 3.7$ as in our simulations, we obtain $\alpha_\mathrm{c} \simeq 0.0070-0.0086$, which is slightly lower than the estimated range in the ICM, $\alpha_\mathrm{c} \simeq 0.01$ (see Sect.~\ref{sec:reference_parameters}). While the uncertainties in $\sigma$, $\phi$ preclude us from making strong statements, these estimates raise the possibility that in significant portions of the ICM the MTI might replaced or supplemented by other sources of turbulence, a scenario that we explore more in detail in Sect.~\ref{sec:external-turbulence}.

\subsection{Differences with flat uniform suppression}

We now summarize our findings on the MTI with whistler suppression and compare to a simple model of uniform suppression of thermal conductivity. As we saw in Sect.~\ref{sec:2d-3d-suppression}, both models show direct relationship between the properties of the turbulence and the level of suppression of thermal diffusivity, whether that is a fixed ratio uniform in space, or a volume-averaged quantity. In this respect, the two models do not differ significantly, and as a result -- given an average estimate of whistler suppression below the Spitzer value in the ICM -- it is in principle possible to determine the resulting level of MTI turbulence. However, care must be taken when extrapolating to the cluster's periphery (or when running simulations with the simpler model of flat uniform suppression) because, as we have seen in Sect.~\ref{sec:ssd_shutdown}, for values of the suppression parameter $\alpha$ above a certain critical threshold the SSD shuts down and the MTI decays. This death-spiral is fundamentally a nonlinear feedback intrinsic to our model of whistler suppression and would not take place with flat uniform suppression. We see this difference most clearly in two aspects: first of all, although with large enough flat suppression the SSD also shuts down, this has no impact on the level of MTI turbulence, which continues unperturbed \citepalias{Perrone2022a}; second, from Fig.~\ref{fig:time_evolution_mti_2d_3d} we saw that a simulation with flat suppression to $40\% \chi_0$ can still sustain an SSD and keep magnetic fields at equipartition, while a whistler-suppressed run with initial volume-averaged suppression of $\simeq 50 \% \chi_0$ (as for the run with $\alpha = 0.12$) does not.
Finally, it is important to keep in mind that the presence of a critical $\alpha$ for the shutdown of the SSD with whistler suppression does not imply the existence of a corresponding threshold value of thermal diffusivity between self-sustaining and death-spiraling MTI. 
Rather, below the critical $\alpha$ with no-net flux the two states (turbulent and dead state) are always available to the system.

\begin{figure*}
        \centering
        \includegraphics[width=1.0\linewidth]{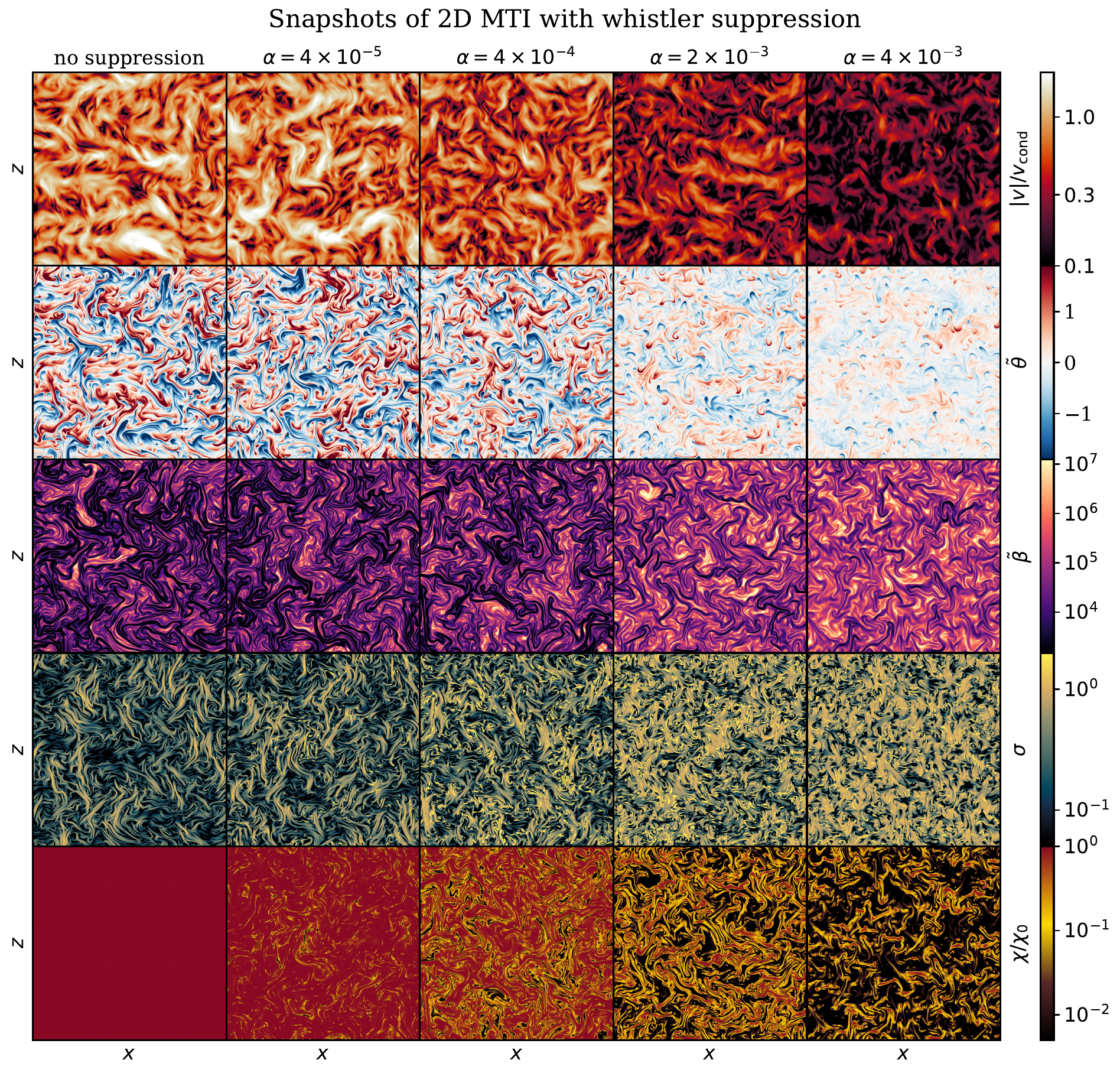}
        \caption{Multiplot of the 2D runs at saturation ordered by increasing suppression of thermal diffusivity (from left to right). From top to bottom: Magnitude of the velocity field, density fluctuation, Boussinesq plasma beta, Boussinesq parallel temperature length, point-wise thermal diffusion suppression fraction. The color bar is in common for each row. A higher suppression parameter $\alpha$ leads to weaker MTI turbulence, weaker magnetic fields (higher $\tilde{\beta}$), less isothermal fieldlines (higher $\sigma$), and lower average thermal conductivity $\chi$.}
        \label{fig:mti2D_final_states}
\end{figure*}

\section{Modifications to the flow with whistler suppression}\label{sec:flow-backreaction}

In our discussion so far we have mostly concerned ourselves with volume-averaged quantities, such as the correlation between the turbulent kinetic energy and the volume-averaged suppression of thermal diffusivity.
However, our model of whistler suppression is a nonlinear function of quantities (such as the modified plasma $\tilde{\beta}$) that can strongly vary in space and across scales, and therefore we expect this variation to be reflected in the suppression of thermal conductivity as well. This is shown in Sect.~\ref{sec:flow-morphology}.
We then take a closer look at the spatial correlations of MTI turbulence and we look at how whistler suppression affects the energy distribution and the MTI energy transfers in spectral space. We find that high-$\beta$ regions are progressively brought out of thermal equilibrium as $\alpha$ increases (Sect.~\ref{sec:probability-distrib}). In Sect.~\ref{sec:spectral-dynamics} instead we observe that -- while whistler suppression reduces the efficiency of MTI turbulent driving -- this effect appears to be mostly limited to the largest turbulent scales.

\subsection{Morphology of the flow and spatial correlations}\label{sec:flow-morphology}

To visualize how the MTI turbulence reacts to increasingly stronger whistler suppression, in Fig.~\ref{fig:mti2D_final_states} we show snapshots of the 2D runs after they have reached saturation (time $t=150$). We focus here on 2D simulations because they allow us to explore a wider range of suppression parameter $\alpha$ than in 3D, where the differences between the surviving runs are less pronounced.
From a visual analysis of the turbulent velocity and temperature field (first and second row) we can immediately see that higher $\alpha$ leads to a progressive weakening of the MTI turbulence.
This is in spite of the fact that the suppression of thermal diffusion (last row) is not uniform across the domain but is highly inhomogeneous and tends to follow the morphology of the magnetic field (third row). We note, however, that even in the two runs with strongest whistler suppression there are still regions with thermal diffusivity higher than $50 \%$ of its reference value, amounting to 11\% and 4\% of the whole domain, respectively. While the isothermality parameter $\sigma$  undergoes a rather dramatic change as $\alpha$ is increased, its role in determining the amount of whistler suppression is less clear.
Finally, from the velocity amplitude and the $\tilde{\beta}$ field (first and third rows), one can see that the largest scale of the turbulence (the integral scale) also decreases with $\alpha$. We observe a similar behavior in our 3D surviving runs.

To better understand the interplay between $\tilde{\beta}$ and $\sigma$, and to assess to what extent the two quantities are spatially correlated or anti-correlated, we employ a variety of diagnostics, including level plots, as well as one- and two-dimensional probability distributions.

\begin{figure}
        \centering
        \includegraphics[width=1.0\columnwidth]{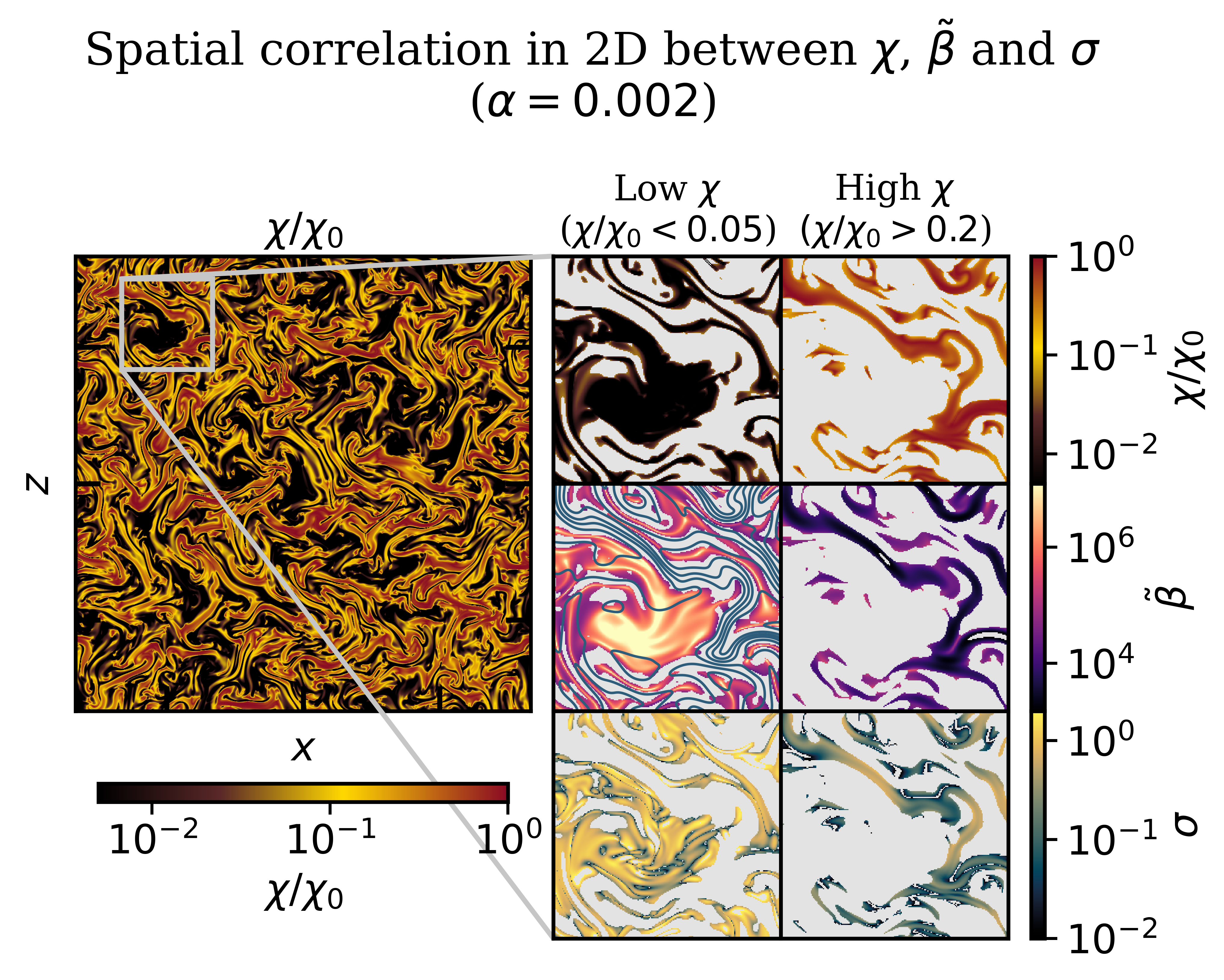}
        \caption{Zoomed-in view of the snapshot of the 2D run with $\alpha = 2 \times 10^{-3}$ at time $t=150$ showing the spatial correlation between $\tilde{\beta}$ and $\sigma$. In the inset we separately show regions with low (left column) and high  (right column) thermal diffusivity by cutting out regions where $\chi / \chi_0$ is above $0.05$ and below $0.2$, respectively. In the second row of the inset, left column, the magnetic field lines are shown in blue. With whistler suppression, regions of high thermal conductivity are shaped as thin bundles near filaments of strong magnetic fields and act in a similar way as  Autobahns for the heat-carrying electrons.}
        \label{fig:mti2D_zoom-in_chi_beta_sigma}
\end{figure}

In Fig.~\ref{fig:mti2D_zoom-in_chi_beta_sigma} we show a zoomed-in view of the suppressed thermal diffusivity of one 2D run ($\alpha = 2 \times 10^{-3}$, from Fig.~\ref{fig:mti2D_final_states}) where we separately show in the inset the regions with  low ($\chi / \chi_0 < 0.05$, left column), and high diffusivity  ($\chi / \chi_0 > 0.2$, right column) and plot the values of $\tilde{\beta}$ and $\sigma$. As we can see, the morphology of the low- and high-diffusivity regions is rather different: while low-$\chi$ regions appear as ``holes'', areas with high-$\chi$ mostly look filamentary. A similar picture can be observed in 3D as well. We relate this behavior to the intermittency of magnetic fields in MHD turbulence, with magnetic field lines (shown in the second row of the inset) concentrated in thin bundles near regions of strong magnetic fields. As a result, with whistler suppression thermal conductivity remains high in these thin bundles (which can be thought of as the ``Autobahns'' of heat-carrying electrons) while it drops significantly in between. Moreover, from Fig.~\ref{fig:mti2D_zoom-in_chi_beta_sigma} we see an interesting interplay emerge between the modified plasma $\tilde{\beta}$ and the isothermality parameter $\sigma$, 
as these quantities appear to be positively correlated: where $\tilde{\beta}$ is high, also $\sigma$ is consistently above $1$, which indicates that the plasma is locally out of isothermality; conversely, in regions with strong magnetic field (low $\tilde{\beta}$) the isothermality parameter $\sigma$ is typically less than unity, and the plasma close to isothermality. This positive correlation, however, is the result of two combined effects, which highlights the nonlinearity of the model: with whistler suppression the thermal diffusivity first drops in regions with high $\tilde{\beta}$, leading to an increase in $\sigma$ as the plasma cannot thermalize as effectively as before. This increase in $\sigma$, however, further reduces the thermal diffusivity (Eq.~\ref{eq:whistler-diffusivity}), and will therefore lead the plasma farther away from isothermality. These effects end up reinforcing each other, and produce the tight correlations shown in Fig.~\ref{fig:mti2D_zoom-in_chi_beta_sigma}.

\begin{figure}
        \centering
        \includegraphics[width=1.0\columnwidth]{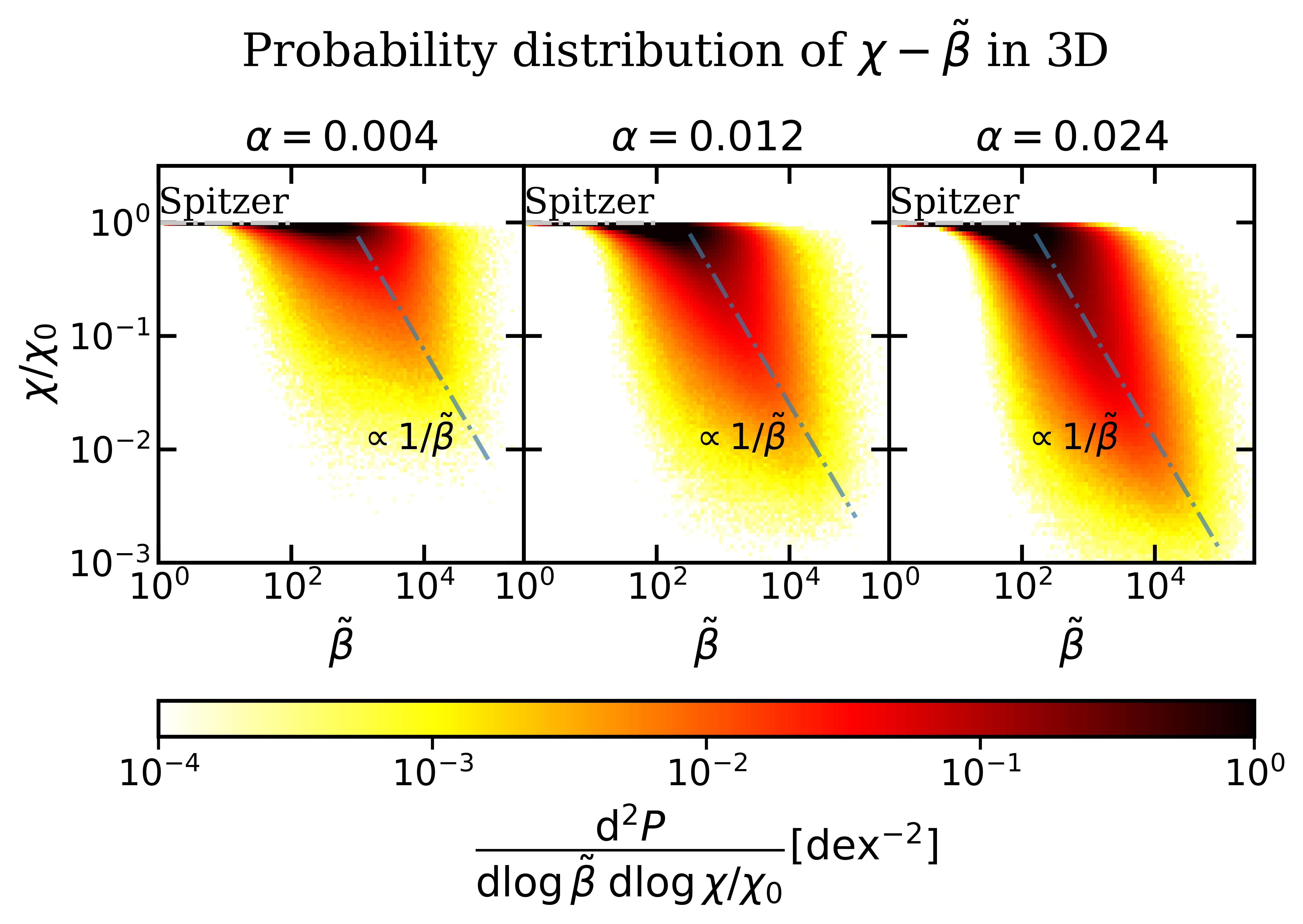}
        \caption{Probability distribution as a function of $\chi/\chi_0$ and $\tilde{\beta}$ in our 3D runs with whistler suppression at saturation. The color bar represents the probability density of finding a value in the infinitesimal interval of size $\mathrm{d}\log (\chi/\chi_0) \, \mathrm{d}\log \tilde{\beta}$. The two dash-dotted lines represent the two limits of Spitzer (in gray)  and whistler-suppressed thermal diffusivity (in blue). By increasing the suppression parameter $\alpha$, larger parts of the domain enter the $1/\tilde{\beta}$ whistler-suppressed regime.
    	}
        \label{fig:mti3D_pdf_chi_beta}
\end{figure}

\begin{figure*}
        \centering
        \includegraphics[width=1.0\linewidth]{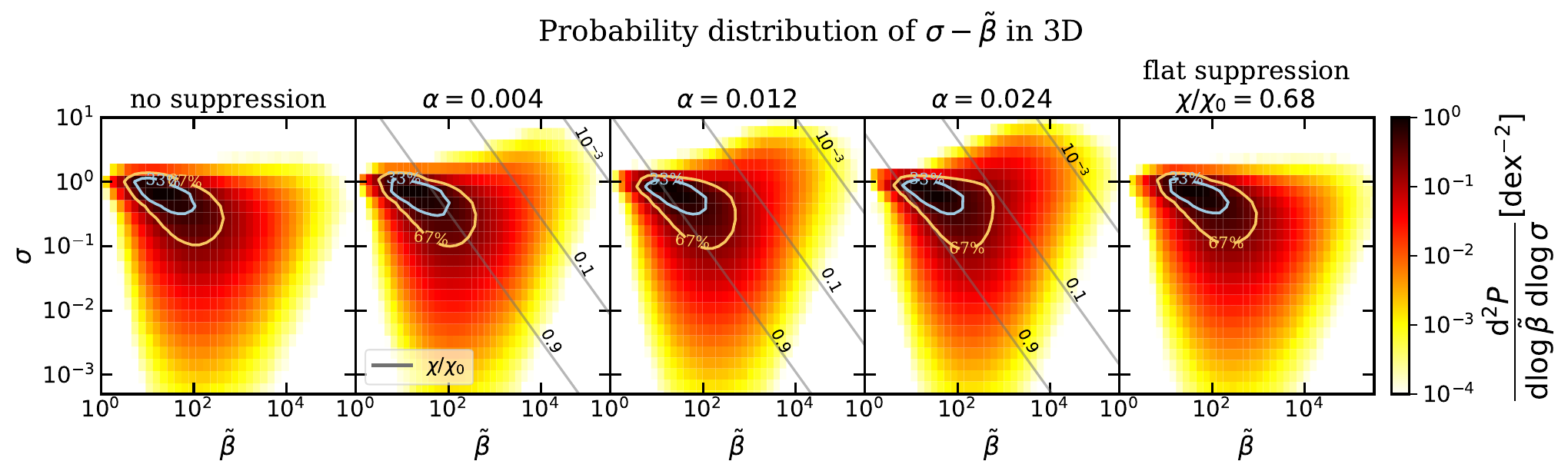}
        \caption{Distribution of the data in the $ \tilde{\beta}$--$\sigma$ plane for our 3D runs at saturation. The color bar represents the probability density of finding a value in the infinitesimal interval of size $\mathrm{d}\log \tilde{\beta} \; \mathrm{d}\log \sigma$. 
    	The cyan and gold curves are isocontours of the probability distribution function that enclose $33\%$ and $67\%$ of the box volume, respectively.
    	The solid gray contour lines in the plots with whistler suppression show the amount of suppression of the thermal diffusivity (to the left of  any given line  are regions with diffusivity larger than the number quoted, in units of $\chi_0$). Due to the intermittent nature of magnetic fields, with whistler-suppression most of the plasma is located in regions with strong magnetic fields (low $\chi$ suppression), while at the same time we note the formation of high-$\beta$ lobes consisting of regions that are not isothermal along field lines. This feature is not captured in the reference simulations with flat or no suppression.}
        \label{fig:mti3D_pdf}
\end{figure*}

\subsection{Probability distributions}\label{sec:probability-distrib}

To complement our visual analysis of spatial correlations, we compute 
a variety probability distribution functions for our 3D runs with whistler suppression.

It is instructive to first look at the distribution of the thermal diffusivities as a function of the modified plasma $\tilde{\beta}$ in the 3D surviving runs (Fig.~\ref{fig:mti3D_pdf_chi_beta}), where we can clearly see the transition between the Spitzer (at low $\tilde{\beta}$) and the whistler-suppressed regime (at high $\tilde{\beta}$) that is encoded in Eq.~\eqref{eq:whistler-diffusivity}. 
The turnover value of $\tilde{\beta}$, where the transition between the two regimes happens (the ``knee''), can be easily estimated by weighing the two terms in the denominator of Eq.~\eqref{eq:whistler-diffusivity}. This yields $\tilde{\beta}_\rmn{turnover} \approx 3 /(\alpha \sigma)$, which shows that it decreases as the suppression parameter $\alpha$ increases. As a result, for higher $\alpha$ a larger proportion of the domain enters the whistler-suppressed regime.
The vertical spread in $\chi / \chi_0$ at fixed $\tilde{\beta}$ is caused by the different values of the 
isothermality parameter $\sigma$.

\begin{figure}
        \centering
        \includegraphics[width=0.9\linewidth]{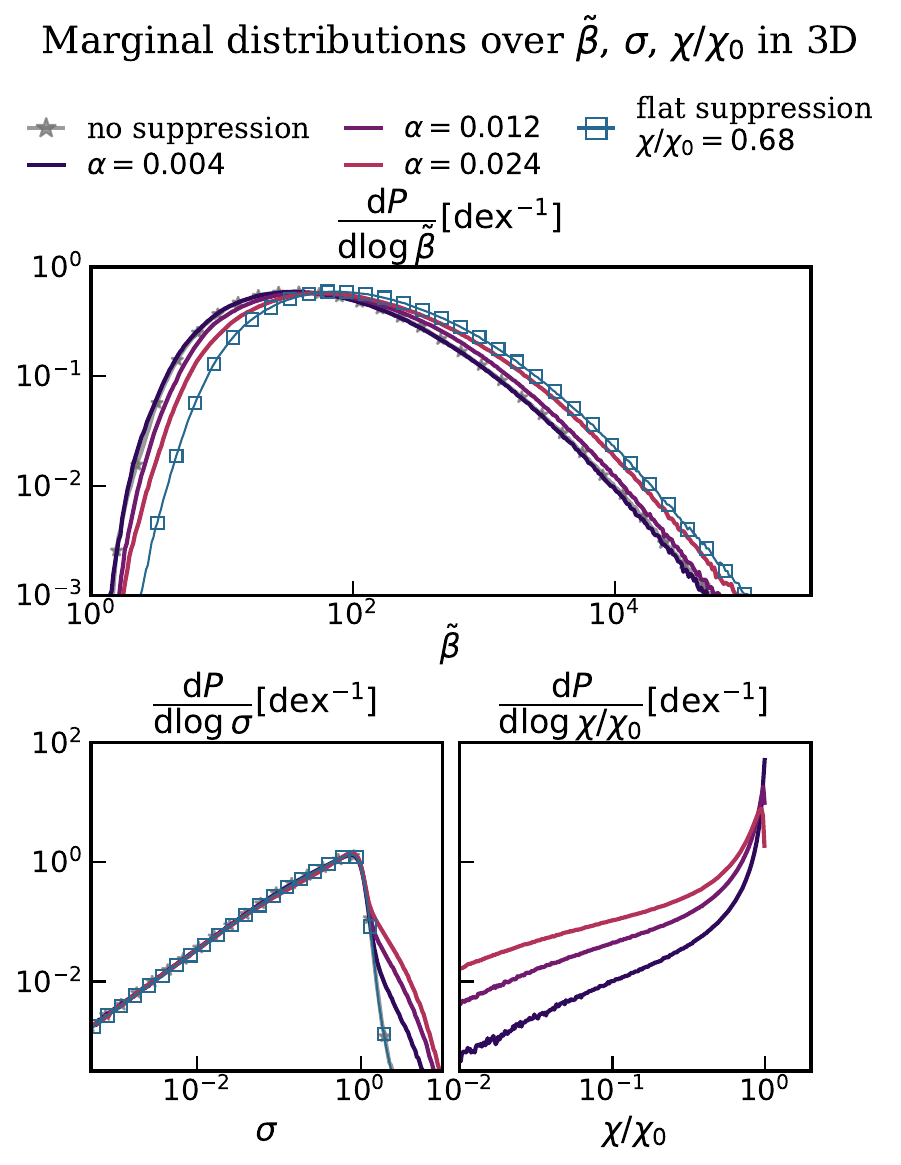}
        \caption{Marginal distributions of the 3D runs with respect to the modified plasma $\tilde{\beta}$ (top), the isothermality parameter $\sigma$ (bottom left), and the effective thermal diffusivity $\chi/\chi_0$ (bottom right). Stronger whistler suppression decreases the overall magnetic energy levels and brings part of the domain out of isothermality.}
        \label{fig:mti3D_pdf_chi_beta_sigma}
\end{figure}

We then show in Fig.~\ref{fig:mti3D_pdf} the two-dimensional probability density of finding a cell with given $(\tilde{\beta},\sigma)$ in our 3D surviving runs, while in Fig.~\ref{fig:mti3D_pdf_chi_beta_sigma} we separately show the marginal distributions along $\chi, \beta$, and $\sigma$ (also in 3D). Without whistler suppression (Fig.~\ref{fig:mti3D_pdf}, leftmost panel), values are distributed in the $\tilde{\beta}$--$\sigma$ plane to resemble a triangle, with the vast majority of the volume ($>95\%$) lying below the $\sigma=1$ line, meaning that the MTI effectively enforces isothermality along magnetic field lines. The values with the highest occurrence are distributed in the oblong-shaped region defined by the isocontours around $\tilde{\beta} \sim 10^2$ and $\sigma \in [10^{-1},10^0]$. Increasing $\alpha$, we note a progressive shift of the high-$\tilde{\beta}$ part of the distribution toward larger values of $\sigma$ (a ``lobe''), with a few data points reaching as far as $\sigma=10$. However, even when $\alpha=0.024$ the fraction of the volume far away from isothermality ($\sigma > 1$) is no more than $9\%$. Interestingly, if we compute the two-dimensional probability density of the local turbulent velocities and suppression of thermal diffusivity, we do not find any positive correlations, contrary to what we observe at the volume-average level.

The fact that only high-$\beta$ regions deviate from isothermality is because it is precisely these regions that are impacted most by whistler suppression when the $\alpha$ parameter is increased (as can be seen by the leftward shift of the $\chi/\chi_0$ contour lines in Fig.~\ref{fig:mti3D_pdf}). When thermal diffusivity is strongly suppressed locally,  thermalization along magnetic field lines stops being efficient and $\sigma$ increases. This behavior is a direct consequence of whistler suppression, and does not take place in simulations with a flat uniform suppression of thermal diffusivity. To show this, we plot in Fig.~\ref{fig:mti3D_pdf} (rightmost panel) the probability distribution of a run with flat suppression where the (spatially uniform) suppression coefficient is chosen to be equal to the volume-averaged suppression factor of the run with $\alpha = 0.024$ (second-last panel). With flat suppression, the distribution of values in the $\tilde{\beta}$--$\sigma$ plane is surprisingly similar to the reference run, except for an overall shift toward larger $\tilde{\beta}$ in order to maintain equipartition with a weaker turbulence, and we observe no lobe out of isothermality.

This picture is further supported by the marginal distributions shown in Fig.~\ref{fig:mti3D_pdf_chi_beta_sigma}, which show the progressive emergence of fat tails for $\sigma >1$ in the one-dimensional distribution (bottom left panel), while for $\sigma < 1$ the distribution is practically unchanged. In the run with flat uniform suppression, instead, the marginal distribution over $\sigma$ is practically identical to that of the reference unsuppressed run. Finally, with whistler suppression the $\tilde{\beta}$-distribution undergoes an overall shift toward larger values (weaker magnetic fields, top panel) in the same way as the case with flat uniform suppression. Nevertheless, we note that for similar values of the volume-averaged diffusivity, in the run with $\alpha = 0.024$ there are more occurrences of regions with strong magnetic fields (low-$\beta$) compared to the run with flat suppression, indicating a higher degree of intermittency. We thus distinguish between a ``local'' and ``global'' response of the turbulence to whistler suppression, where the local response involves the inhibition of isothermality in those regions where $\tilde{\beta}$ is large, while the global response is the overall decrease in turbulence strength and magnetic energy in the system.

\subsection{Dynamics in Spectral Space}\label{sec:spectral-dynamics}

\begin{figure*}
        \begin{subfigure}{.5\linewidth}
            \centering
            \includegraphics[width=0.9\columnwidth]{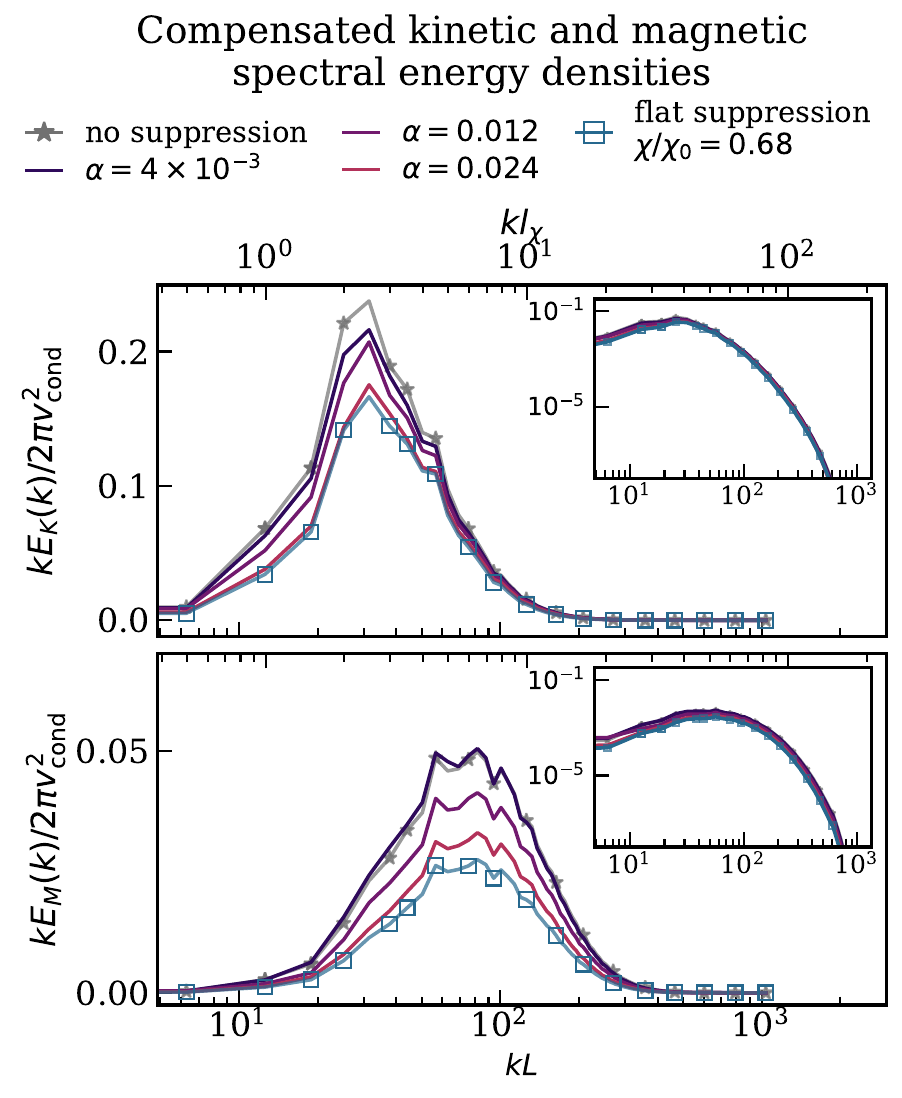}
            \caption{}
            \label{fig:mti3D_spectra_comparison}
        \end{subfigure}%
        \begin{subfigure}{.5\linewidth}
            \centering
            \includegraphics[width=0.9\columnwidth]{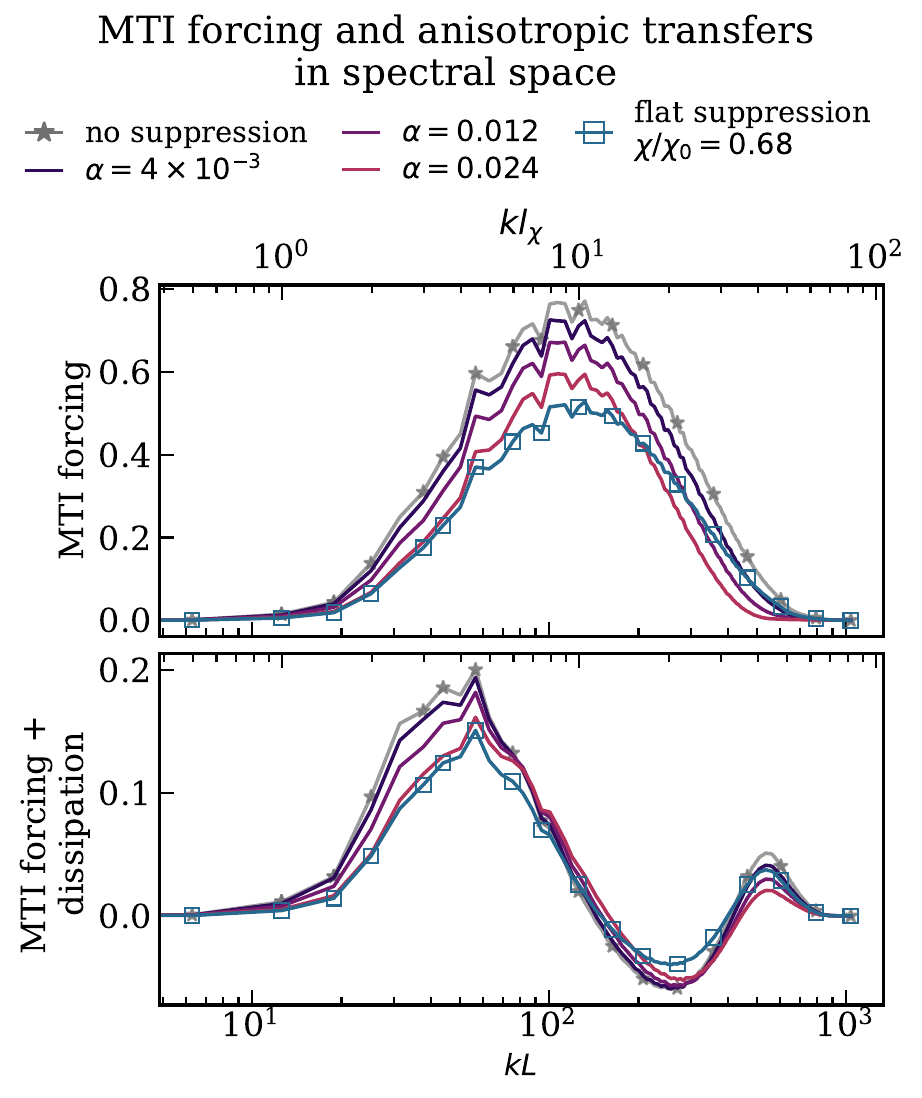}
            \caption{}
            \label{fig:mti3D_forcing_diss_comparison}
        \end{subfigure}%
        \caption{Spectral energy densities and spectral energy fluxes in our 3D runs at saturation, averaged over hundreds of dynamical times. Left: Compensated spectral energy densities (i.e., multiplied by the wave number $k$) of the kinetic (top) and magnetic fluctuations (bottom). The $y$-axis is in linear scale to highlight the differences between the runs. For comparison, we show in the insets the kinetic and magnetic (uncompensated) power spectra in log-log scale. Right: compensated MTI energy injection (top) and balance between MTI injection and anisotropic thermal dissipation (bottom) in spectral space in 3D simulations. 
        Despite the reduction in MTI energy injection due to increasingly strong whistler suppression, the final balance between injection and dissipation remains mostly unchanged over a wide range of scales below the conduction length $l_{\chi}$.
        }
        \label{fig:mti3D_spectral_analysis}
\end{figure*}

We now look at the impact of whistler suppression on thermal conductivity across scales, using spectral energy densities (power spectra) and transfer terms in spectral space.
In Fig.~\ref{fig:mti3D_spectra_comparison} we show the kinetic and magnetic power spectra as a function of $k$, the wavenumber, for the reference MTI run and the three surviving 3D runs with whistler suppression. To aid visual inspection, we compensate the spectra multiplying by $k$ and use a linear scale in the $y$-axis. For comparison, the same uncompensated spectral energy densities in log-log scale can be found in the insets of Fig.~\ref{fig:mti3D_spectra_comparison}. Looking at the kinetic power spectrum (top panel), we find that increasing $\alpha$ mostly affects the energy contained at the largest scales, while the small scales are hardly affected. This behavior is reminiscent of the MTI without whistler suppression, where decreasing the (spatially uniform) thermal diffusivity leads to a decrease of energy at the largest scales, with the integral scale moving to larger $k$ \citepalias{Perrone2022a}. We note, however, that with whistler suppression this shift is hardly noticeable given that the runs shown only have a limited amount of suppression of thermal conductivity ($\langle \chi / \chi_0 \rangle \gtrsim 68\%$). Looking at the magnetic power spectrum, on the other hand, we find that the decrease in energy is more evenly distributed across wavenumbers. 

In Fig.~\ref{fig:mti3D_spectra_comparison} we also show the spectra of the run with flat uniform suppression and $\chi / \chi_0 = 68\%$, and see that they align rather well with the equivalent whistler-suppressed run with $\alpha = 0.024$ (particularly so in the kinetic power spectrum), suggesting a similar dynamics in spectral space. We can get additional insight by looking at the transfers of energy in spectral space. We focus on two key processes of the MTI, namely the injection of energy from the background temperature gradient by the anisotropic conductivity and the net injection plus anisotropic dissipation, which is a more meaningful measure of how quickly energy is eventually added or removed to the system. 
The behavior of these two terms (shown in Fig.~\ref{fig:mti3D_forcing_diss_comparison}, top and bottom panels, respectively) is particularly revealing: while they individually show a clear downward trend as $\alpha$ is increased (meaning that the energy injection and dissipation become less efficient both at large and small scales), their sum is instead rather constant, except at the largest scales comparable to the conduction length $l_{\chi}$. 
If we now look at the run with flat uniform suppression and compare it with the run with $\alpha = 0.024$ (that has the same volume-averaged thermal diffusivity), we find that the energy injection efficiency looks clearly different: at intermediate scales ($kL \lesssim 100$) MTI forcing is more efficient with whistler suppression than with flat uniform suppression, while at small scales the situation is reversed. At the largest scales the two are very similar. Interestingly, these differences again mostly vanish once we look at the net effect of injection plus dissipation.

\subsection{Differences with flat uniform suppression}

In the last sections we carried out a detailed study of the statistical properties of MTI with whistler suppression, and looked at how the spectral dynamics is affected. The comparison with a model of flat uniform suppression helped us identify some key features of whistler suppression, which we classified as ``local'' (the bringing of high-$\tilde{\beta}$ regions out of isothermality) or ``global'' (the overall decrease of magnetic field strength). Both of these outcomes, tightly related to magnetic field intermittency, influence the level of whistler suppression itself, and therein lies the nonlinearity of the process. 
The differences in the spectral space are more subtle: despite the clear change in the MTI forcing efficiency, the net contribution of injection and dissipation due to anisotropic diffusion shows striking similarities with simulations with flat suppression, which helps explain why the kinetic and magnetic power spectra are also very much alike.

\begin{figure}
        \centering
        \includegraphics[width=1.0\linewidth]{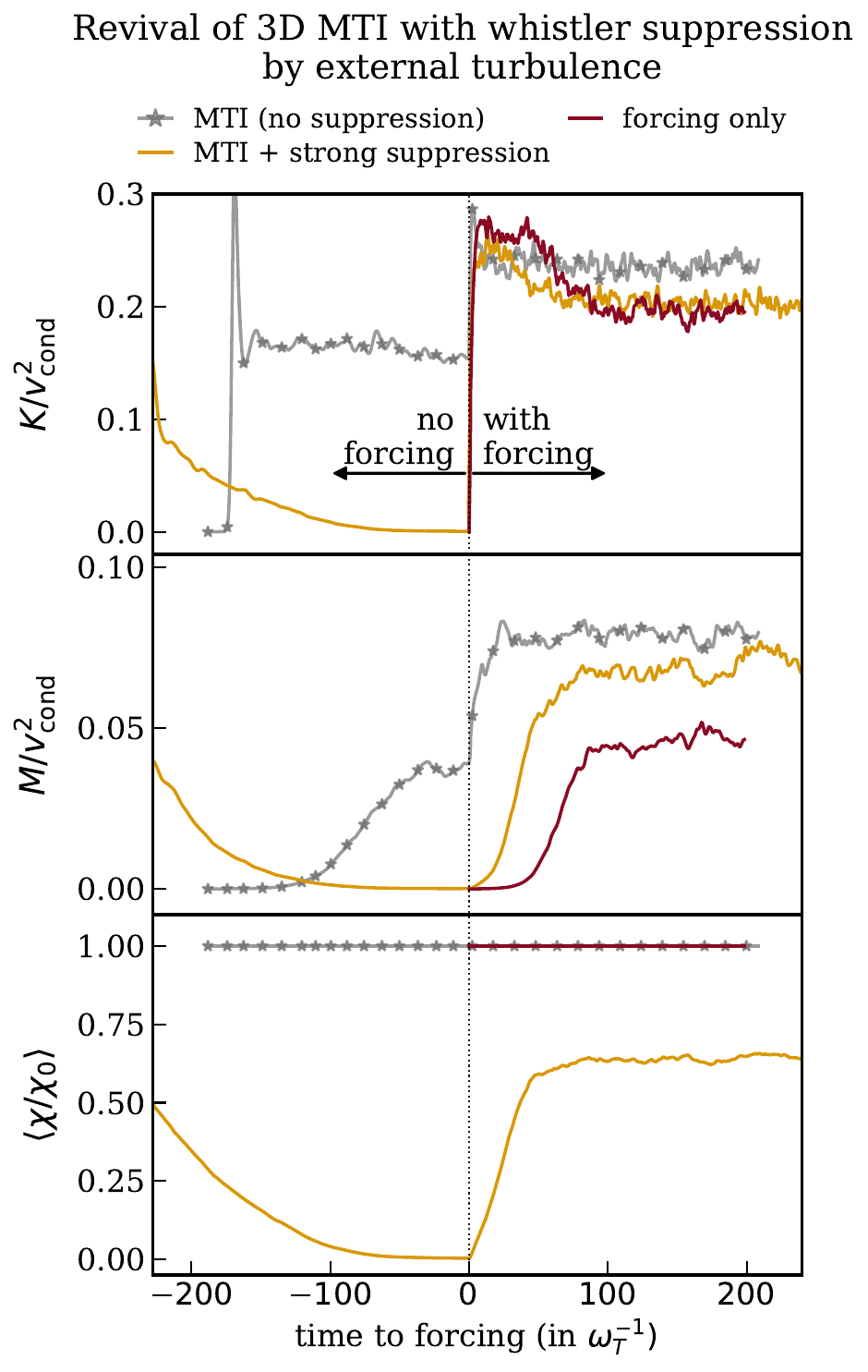}
        \caption{Time evolution of volume-averaged kinetic, magnetic energy and average suppression of thermal diffusivity in 3D with and without external forcing. The zero point on the $x$-axis denotes the time when external forcing was activated in the runs. External forcing drives its own SSD and in so doing revives previously dead MTI simulations with strong whistler suppression.}
        \label{fig:mti3D_forc_time_evolution}
\end{figure}

\section{Three-dimensional MTI turbulence with external forcing and whistler suppression}\label{sec:external-turbulence}

\begin{figure*}
        \centering
        \includegraphics[width=1.0\linewidth]{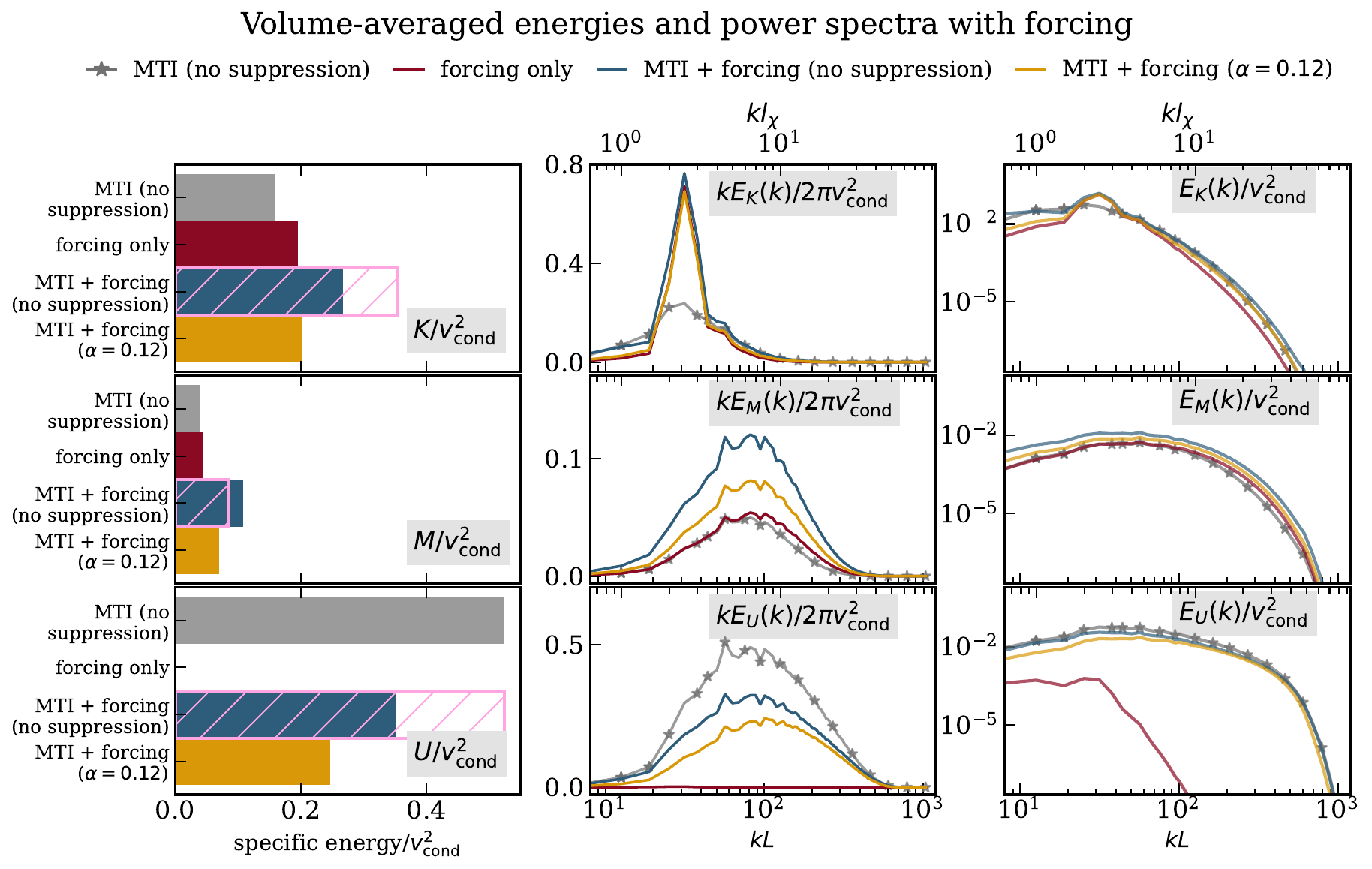}
        \caption{Volume-averaged and spectral energy densities for the 3D runs with and without external forcing at saturation. From left to right: Bar plot of the volume-averaged turbulent energies, compensated (i.e., multiplied by $k$), and uncompensated spectral energy densities. 
    	The color legend is in common between the bar plots and the power spectra. The pink hashed bars in the left column mark the sum of the energy levels of the MTI-only and forcing-only runs. The peak in the kinetic energy near $l_{\chi}$ indicates the wave numbers excited by the external forcing. The turbulent driving of the MTI and the external forcing do not add up exactly for the kinetic energy, but nearly so for the magnetic energy. Moreover, the MTI excites velocity fluctuations over a wide range of scales below $l_{\chi}$.}
        \label{fig:mti3D_forc_spectra}
\end{figure*}

We now turn our attention on the impact of external sources of turbulence on MTI turbulence with whistler suppression. As we show, external forcing can actually have a beneficial effect, reviving simulations that had entered the death-spiral because the suppression parameter $\alpha$ is above the critical threshold for the SSD (Sect.~\ref{sec:revival-forcing}). We then look more in detail at the interplay between MTI and externally driven turbulence, trying to distinguish signatures of the MTI (with and without whistler suppression) in simulations with external forcing, which may have implications for observations.
We find that the presence of the MTI is clearly recognizable in spectral space, where we observe an excess of energy in the turbulent fluctuations at small scales, compared to a forcing-only simulation with similar volume-averaged kinetic and magnetic energies (Sect.~\ref{sec:signature-mti}). Interestingly, we also find that the inclusion of externally driven turbulence does not significantly affect the statistical properties of whistler suppression, except for an overall increase of the magnetic energy due to the additional SSD effect (Sect.~\ref{sec:forcing-statistics}).

\subsection{Revival of the dead MTI with external forcing}\label{sec:revival-forcing}

In Fig.~\ref{fig:mti3D_forc_time_evolution} we show the time evolution of the volume averages of kinetic energy, magnetic energy and thermal conductivity for four representative 3D runs (also discussed in \citetalias{Perrone2023}): the MTI reference run (without whistler suppression), a forcing-only run with isotropic thermal conductivity without the MTI, a run with both MTI and external forcing (with no whistler suppression), and finally a run with MTI, external forcing and whistler suppression ($\alpha = 0.12$) which was firmly in the death-spiral regime (see Fig.~\ref{fig:time_evolution_mti_2d_3d}). 
We choose the level of external forcing so as to drive turbulence at saturation of similar amplitude and at similar scale as the MTI reference run without whistler suppression (gray starred line). Our choice is supported by previous estimates that the strength of MTI-driven turbulence might be roughly comparable to the observed levels in the periphery of galaxy clusters \citepalias{Perrone2022a}. For more details on the forcing mechanism refer to Sect.~\ref{sec:numerics-external-forcing}. 

By construction, the forcing-only run (solid red line) has saturated kinetic energy levels comparable to those of the MTI (no suppression) reference run before forcing ($0.19$ vs.\ $0.16$, in units of $\varv_{\rmn{cond}}^2$), and similar magnetic energy ($0.045$ vs.\ $0.040$). Due to the absence of the MTI, on the other hand, in the forcing-only run the potential energy is substantially lower ($1.3\times 10^{-3}$ vs.\ $0.51$, not shown). Combining (unsuppressed) MTI and external forcing leads unsurprisingly to larger levels of kinetic and magnetic energies than either the MTI-only or forcing-only runs, although the external and the MTI turbulent kinetic energies do not exactly sum up
\citep[contrary to what was observed by][]{Parrish2012}. Interestingly, the magnetic energy at saturation appears to be closer to the sum of the MTI-only and forcing-only runs. 

The most remarkable consequence of adding external forcing, however, can be seen in the run with strong whistler suppression ($\alpha=0.12$, solid orange line), where the MTI turbulence was previously decaying: after switching on external forcing, both turbulent kinetic and magnetic energies quickly increase and overshoot the levels of the forcing-only run (first and second rows of Fig.~\ref{fig:mti3D_forc_time_evolution}), particularly so for the magnetic energy. Similarly, the volume-averaged thermal diffusivity returns to very high levels, around $67\%$ of the unsuppressed Spitzer value (third row). 
We interpret this evidence as a clear sign that the MTI has been successfully revived by the addition of external forcing, and now contributes to the observed turbulent levels \citepalias[this conclusion is also supported by the positive sign of the buoyancy power, indicating instability, see][]{Perrone2023}. The reason for this phenomenon is that the externally driven turbulence can also drive its own SSD and, contrary to what we examined in Sect.~\ref{sec:ssd_shutdown}, this process is not affected by how much suppression of thermal diffusivity there is. Since turbulent magnetic fields are now propped up by the external forcing, whistler suppression of thermal diffusivity cannot go below a certain value (determined by the equipartition value of $\tilde{\beta}$ with the external turbulence). If the residual thermal diffusivity is high enough, the MTI has a chance to make a comeback and do its work on top of the external turbulence, which is indeed what we observe in Fig.~\ref{fig:mti3D_forc_time_evolution}. Fixing the external forcing, the final turbulent levels of the MTI with both whistler suppression and external forcing will depend on the exact value of the suppression parameter, $\alpha$. However, we can realistically expect that they will span the range between the turbulent levels of the unsuppressed MTI run with forcing (in the limit of low $\alpha$), and those of the forcing-only run without the MTI (if $\alpha$ is so high that the residual thermal diffusivity does not allow the MTI to grow anymore).

This behavior is particularly interesting because it constitutes a first example of a positive interplay between external forcing and the MTI with whistler suppression, which is clearly nontrivial and need not be entirely inimical. 
We can speculate that two limiting regimes may exist: if the external turbulence is too weak, then the thermal diffusivity cannot be boosted to a sufficiently large value to allow the MTI to become again unstable, thus no revival could take place; if the external turbulence is too strong, however, the MTI can be boosted only up to a certain extent ($\chi / \chi_0$ cannot grow larger than 1), and the external turbulence may then again dominate the turbulence so that the MTI contribution is negligible \citep[although this should be thought of as a scale-dependent statement,][]{McCourt2011}. Therefore it appears that a ``Goldilocks'' scenario may arise when the external forcing and the MTI can both operate at comparable strength, which is precisely the regime we studied and that we consider realistic for the periphery of galaxy clusters. In order to make conclusive statements, however, a more detailed exploration of the different forcing amplitudes and driving scales is required.

\subsection{Signature of the MTI in forced-turbulence simulations}\label{sec:signature-mti}

We now look more in detail at the energy distribution in spectral space  at saturation of our runs with forcing. Despite the similarities between the volume-averaged energies of the forcing-only and MTI-only runs, 
their power spectra differ noticeably. In Fig.~\ref{fig:mti3D_forc_spectra} we plot the kinetic, magnetic and potential power spectra (from top to bottom) both in log-log scale (right column) and in linear scale after multiplying by the wave number $k$, to highlight the different features (middle column). As we can see from Fig.~\ref{fig:mti3D_forc_spectra} (top row), the kinetic power spectra of the unsuppressed MTI-only run presents stronger turbulence at all wavelengths below and above the forcing scale compared to the forcing-only run.
This result underlines the fact that the energy injection provided by the MTI works very differently from external forcing: the MTI actually drives the turbulence over a wider range of scales below the conduction length $l_{\chi}$, whereas in our forcing model all the energy is dumped in a small set of wavelengths (which results in a very visible spike near $l_{\chi}$ in the kinetic power spectrum). 
Because of the wide-range injection by the MTI, identifying a proper inertial range -- defined by the absence of both injection and dissipation --
becomes problematic, and one should expect deviations from the classic Kolmogorov scaling.
Interestingly, we note that the magnetic power spectra are much less affected by the change in the driving, with the MTI-only and forcing-only lines closely tracing each other over a wide range of scales (middle panel).

Combining the (unsuppressed) MTI and external forcing we observe stronger velocity fluctuations at all scales \citep[this is consistent with earlier results by][]{McCourt2011} whereas the magnetic power spectrum appears to be boosted by a multiplicative factor, while its shape remains relatively unchanged. Finally, an intermediate behavior is observed in the simulation with whistler suppression and forcing, where the kinetic and magnetic energy levels are below the run with (unsuppressed) MTI and external forcing, but above those in either the MTI-only or forcing-only runs. Interestingly, this trend is reversed in the potential energy power spectra (bottom panel), where the largest thermal fluctuations are achieved in the MTI-only run without whistler suppression, while both external forcing and whistler suppression lead to a decrease in their amplitude, particularly at large scales.

\begin{figure}
        \centering
        \includegraphics[width=\linewidth]{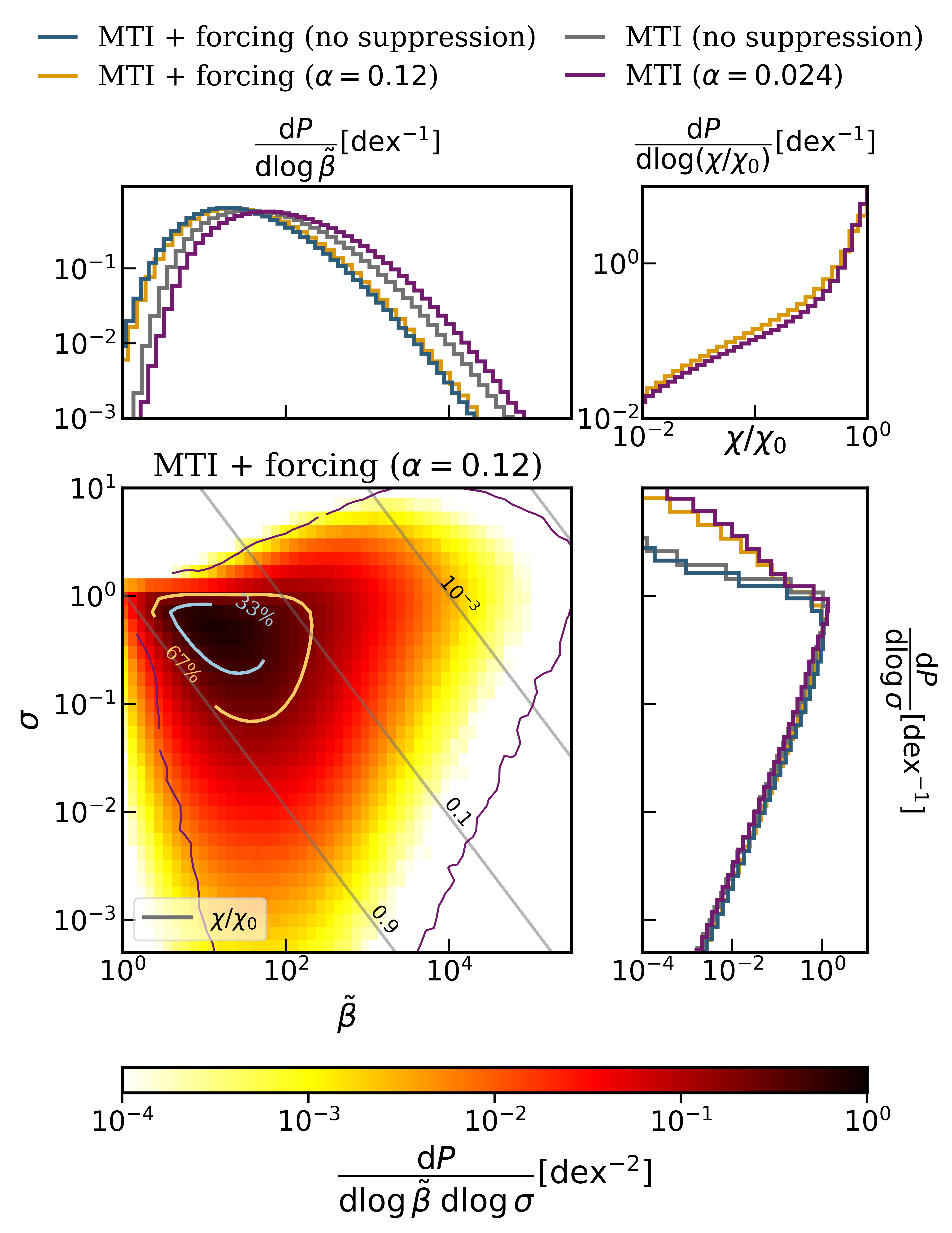}
        \caption{Probability densities and marginal distributions with and without external forcing, and with and without whistler suppression. The overall magnetic field strength mainly depends on whether external forcing is included, while the degree of isothermality is almost exclusively affected by whistler suppression. From bottom left clockwise: Two-dimensional probability density as a function of $\tilde{\beta}$ and $\sigma$ for the MTI run with whistler suppression $\alpha = 0.12$ and external forcing, marginal distribution over $\tilde{\beta}$, marginal distribution over $\chi/\chi_0$, and marginal distribution over the isothermality parameter $\sigma$. In the two-dimensional plot the solid purple line denotes the isocontour of $10^{-4}$ for the MTI run without forcing and $\alpha=0.024$, which shows approximately the same values of the volume-averaged $\chi/\chi_0$.}
        \label{fig:mti3d_forc_pdf_comparison}
\end{figure}

\subsection{Statistics of whistler suppression unaffected by external forcing}\label{sec:forcing-statistics}

In addition to volume-averaged and spectral diagnostics, we are also interested in how the statistics of MTI turbulence with whistler suppression are affected by the external forcing. In Fig.~\ref{fig:mti3d_forc_pdf_comparison} we plot the two-dimensional probability density as a function of $\tilde{\beta}$ and $\sigma$ for the MTI run with strong whistler suppression ($\alpha=0.12$) and external forcing, together with the marginal distributions over $\tilde{\beta}$, $\sigma$, and $\chi/\chi_0$. We note that the addition of external forcing does not significantly affect the probability distribution, which is qualitatively very similar to the runs without forcing analyzed in Sect.~\ref{sec:probability-distrib}: the values are distributed within this roughly triangular shape, with a prominent lobe at high-$\tilde{\beta}$ representing portions of the domain out of isothermality, whereas the majority of the volume is clustered in the region with $\sigma \in [10^{-1},10^0]$ and $\tilde{\beta} \sim 10^1-10^2$. 

In the bottom left plot of Fig.~\ref{fig:mti3d_forc_pdf_comparison} we also show the isocontour of the two-dimensional probability (level $10^{-4}$, solid purple line) of the run with MTI without forcing and $\alpha=0.024$: this run has similar volume-averaged suppression $\langle \chi/\chi_0 \rangle$ as the MTI run with strong whistler suppression ($\alpha=0.12$) and external forcing ($68\%$ vs.\ $64\%$, respectively). Thus, it offers us the possibility to compare two very different scenarios: one with weak whistler suppression and no external forcing, while the other with strong whistler suppression and forcing, but in the end with similar volume-averaged thermal diffusivities. Interestingly, the main difference between the two runs is an overall shift of the distribution toward lower $\tilde{\beta}$ (higher magnetic fields) when external forcing is included.

For a more quantitative comparison, we can look at the marginal distributions of Fig.~\ref{fig:mti3d_forc_pdf_comparison}, where we compare two runs with MTI and external forcing ($\alpha=0$ and $\alpha=0.12$) with two MTI-only runs ($\alpha=0$ and $\alpha=0.024$). 
From the marginal distribution over $\tilde{\beta}$ (top left), we note again how stronger levels of turbulence (due to external turbulence or lower $\alpha$) result in an overall shift of the probability distribution toward lower values of $\tilde{\beta}$, while the shape remains relatively unchanged. This behavior is similar to what we observed in Fig.~\ref{fig:mti3D_pdf_chi_beta_sigma}, and is due to a stronger driving of the SSD. 
We distinguish a split in the marginal distribution over the isothermality parameter $\sigma$ (bottom right)  between the runs with and without whistler suppression. In the runs without whistler suppression, the distributions fall off rapidly for values above $\sigma = 1$, and in the runs with whistler suppression, the distributions develop fat tails for $\sigma >1$. This behavior suggests that the level of isothermality enforced by the MTI is not influenced by external forcing, and that it only depends on the effective thermal diffusivity. Further support to this picture is given in the top right panel, where we plot the marginal distribution of the two runs with whistler suppression with respect to $\chi/\chi_0$: there we see confirmation that not only the volume-averaged $\langle \chi/\chi_0 \rangle$ between the two runs is similar, but also the frequency with which the individual values occur.

In conclusion, there seems to be a level of degeneracy between the suppression parameter $\alpha$ and the magnetic field strength (set by total turbulence levels in MTI simulations with external forcing) where external forcing can effectively compensate for larger values of $\alpha$, and produce similar levels of thermal diffusivity. This similarity is manifested in certain aspects (as the distribution over the isothermality parameter $\sigma$), but not others: with external forcing, the total magnetic energy is higher, whereas the potential energy and the buoyancy power (the rate at which energy is added by the buoyancy force) are both lower ($0.25$ vs.\ $0.43$ and $0.002$ vs.\ $0.007$, in units of $\varv_{\rmn{cond}}^2$ and $\epsilon_0$, respectively).

\section{Discussion}\label{sec:discussion}

\subsection{Can we model the effect of whistler suppression on the MTI as a flat decrease in thermal conductivity?}
We pick up the threads of these last sections and summarize our findings on whether it is appropriate or not to model whistler suppression numerically as a flat decrease in thermal conductivity, and what physics one might miss by doing so. We stress that this summary is necessarily focused on the differences seen in Boussinesq MTI-driven turbulence and that we cannot exclude significant differences with more sophisticated physical models that take into account (e.g., radiative physics, galaxy formation).

In Sect.~\ref{sec:2d-3d-suppression} we observed that for modest suppression below the Spitzer value of thermal diffusivity, whistler suppression behaves very similarly to a flat uniform suppression in terms of the turbulent levels and other volume-averaged quantities. This is consistent with the theoretical model proposed in \citetalias{Perrone2022} where the turbulent root-mean-square velocities are determined by the MTI energy injection rate and the largest scale excited by the turbulence, both of which depend on $\chi$. A similar picture emerges from looking at the power spectra in Sect.~\ref{sec:spectral-dynamics}, which look remarkably similar to those computed for simulations with a flat uniform suppression.
Despite some noticeable differences in the MTI energy injection and anisotropic dissipation with whistler suppression, these differences are averaged out such that the net energy injection rate in wave number space is very close to the uniform suppression model. 
However, by inspecting the spatial distribution of turbulent quantities, we saw that with whistler suppression heat is transported predominantly along thin bundles of strong magnetic fields, while regions with high plasma $\beta$ can be locally brought out of isothermality. Both these features would not be observed with a model of flat uniform suppression.

\begin{figure}
        \centering
        \includegraphics[trim={5cm 3cm 6cm 2cm},clip,width=1.0\linewidth]{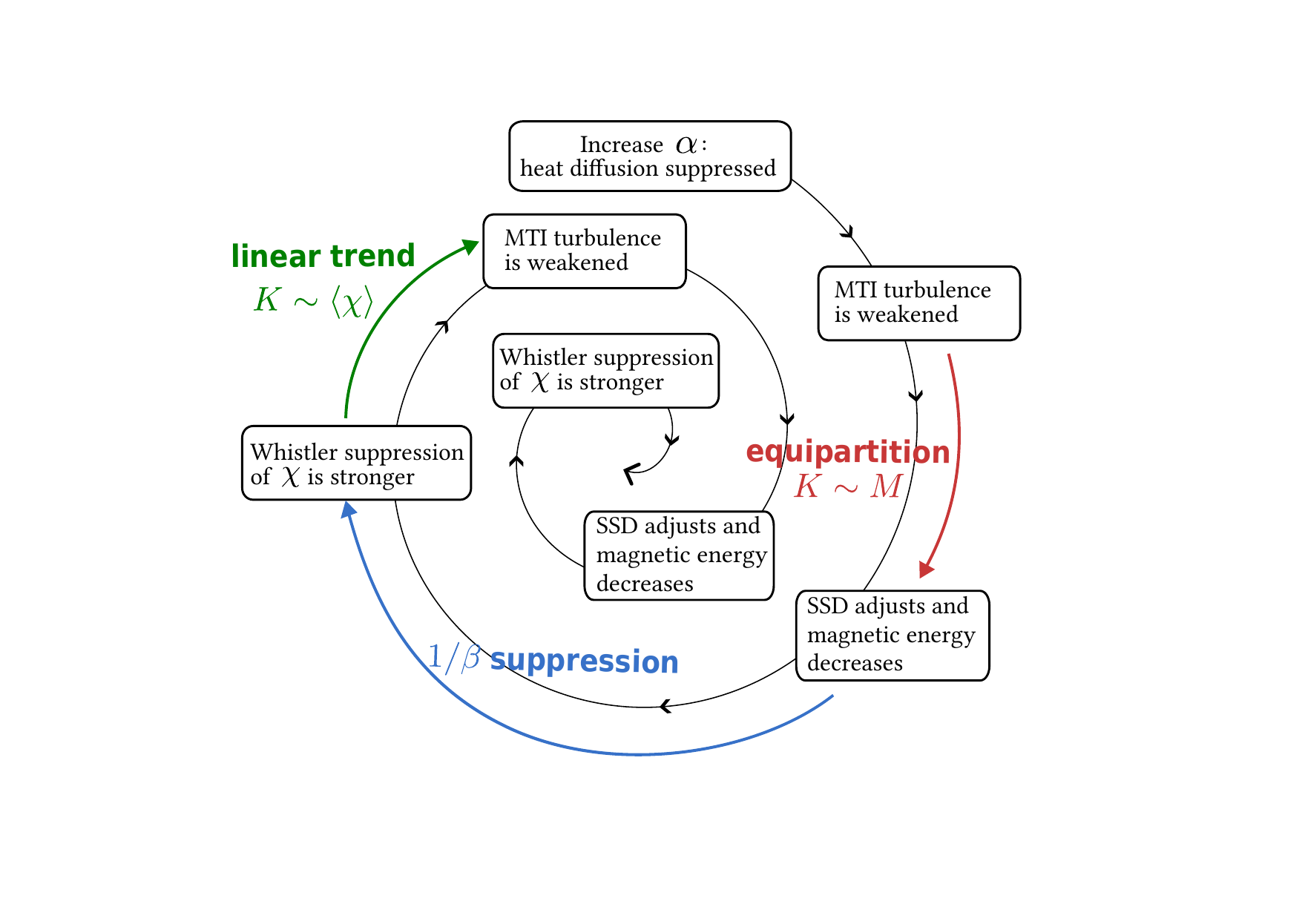}
        \caption{Representation of the death-spiral mechanism with whistler suppression. When the suppression parameter $\alpha$ is increased beyond a critical value, the turbulence weakens and the magnetic field strength decreases to maintain equipartition. As a result, whistler suppression gets stronger and the turbulence weakens further.}
        \label{fig:conclusion-death-spiral}
\end{figure}

The most striking piece of new physics introduced by whistler suppression, however, is that when the level of suppression becomes too large the SSD is shut down and the system enters into what we called the death-spiral. An illustration of the death-spiral can be found in Fig.~\ref{fig:conclusion-death-spiral} and can be summarized as follows: starting from a turbulent equilibrium state, when the suppression parameter $\alpha$ increases beyond a critical value (either by tuning it manually as in our local MTI simulations, or because the physical conditions in the environment change) the volume-averaged thermal diffusivity decreases and so does the strength of the MTI-driven turbulence in order to keep the linear relationship between $\langle \chi \rangle$ and the turbulent kinetic energy. Once the strength of the turbulence becomes weaker, the SSD follows suit and the total magnetic energy adjusts to the new (lower) equipartition-value. However (and here lies the difference between whistler suppression and a flat uniform suppression), the lower magnetic energy directly influences the amount of suppression of thermal diffusivity through the $1/\beta$ dependence, which in turn leads to weaker turbulence and so forth. Even though the suppression parameter $\alpha$ appears to be near the critical value for the shutdown of the SSD in the cluster periphery (Sect.~\ref{sec:ssd_shutdown}), different regions of the cluster periphery may have very different conditions (e.g., they could experience a merger that temporarily stirs the ICM so that the pressure scale height locally increases, or have locally weaker magnetic fields or stronger temperature gradients), and therefore a shutdown of the MTI cannot be expected a priori everywhere.
By using a flat suppression model this behavior would be missed entirely, which underscores the importance of adopting well-motivated subgrid models. 

\subsection{Limitations of this study}

This work comes with several limitations which we can broadly categorize as (i) intrinsic to the specific subgrid model chosen for whistler suppression, and (ii) pertaining the application of our results to galaxy clusters. In the first camp, the main unknown is certainly to what extent the closure for thermal conductivity that we adopted represents an accurate description of the effect of whistler waves on electron transport in the bulk of the ICM. There are a number of reasons why the model in Eq.~\eqref{eq:whistler_suppression_approx} can only constitute a partial approximation of the true thermal conductivity in galaxy clusters.
First of all, the whistler-dominated regime for the heat flux (the $1/\beta$ suppression) has been convincingly shown only in kinetic simulations with very strong temperature gradients \citep[on scales not much longer than electron gyroradii; see, e.g.,][]{Roberg-Clark2016,Roberg-Clark2018,Komarov2018}, which are necessary to drive the whistler instability. Such small-scale temperature gradients are unlikely to be long-lived in galaxy clusters, with typical mean-free paths $14$ orders of magnitude larger than the electron gyroradii. Other ways of exciting whistler waves in the ICM exist \citep[e.g., they could appear as side-products of mirror modes, driven by the anisotropic pressure rather than a temperature gradient,][]{Ley2023}, but if so the appropriateness of the closure introduced in Eq.~\eqref{eq:whistler-dominated-hflux} for the heat flux becomes less clear.
Other sources of uncertainty lie in the fact that the interpolation formula in Eq.~\eqref{eq:whistler_suppression_approx} implicitly assumes a hierarchy of temporal and spatial scales between the slow, large-scale fluid motions (such as those induced by the MTI) and the fast, micro-scale whistler instability, with the latter instantaneously adjusting to the changing conditions of the plasma, without interference by the large-scale turbulence. Even though this is plausible on the fluid scales that we simulated in this study, in theory the MTI (or its kinetic counterpart) can be excited at the same electron scales where the whistler instability develops \citep{Xu2016}. As a result, the interplay between the two instabilities is likely to be more complex than anticipated. Finally, there is currently no consensus on what the combined impact of whistler, mirror and firehose instabilities on the thermal conductivity could look like, and therefore these additional effects were also neglected in our study.

While we believe that our choice to perform small-scale simulations using the Boussinesq approximation has proved very fruitful, as it allowed us to gain a solid understanding of the nonlinear interplay between the MTI and whistler suppression of thermal diffusivity, which would have been hard to obtain otherwise, it comes with some caveats and inevitably limits the applicability of our findings to model the large-scale dynamics in the periphery of galaxy clusters. 
First of all, one possible limitation of our 3D MTI simulations is that they have been constructed in a way that is not entirely self-consistent, since we spawned our ensemble of whistler-suppressed MTI runs from an unsuppressed reference MTI run that had reached saturation. By doing so, a possible concern is that our results might be partial to these specific initial conditions. However, we believe that ours was the best option available within our numerical and physical setup, and that the main findings of this work are largely unaffected by it. Had we chosen to initialize our 3D runs with an initially uniform horizontal magnetic field (as is typical for MTI simulations), we would have introduced an unavoidable degree of arbitrariness, in that the choice of the initial value of $B_0$ would have directly impacted the amount of whistler suppression and determined a priori whether the MTI could be linearly destabilized or not. 
Starting from a pre-existing turbulent field, on the other hand, offers the advantage of having realistic equipartition-level magnetic fields that are self-consistent (i.e., not uniform) and strong enough to keep thermal conductivity at a reasonable value based on our estimates of whistler suppression in the periphery of galaxy clusters. Admittedly, such initial conditions need not be generated through our reference MTI simulation without whistler suppression \citep[a forced-turbulence simulation without the MTI would probably have worked as well, since the MTI can also be nonlinearly destabilized, see][]{McCourt2011,Perrone2022a}, 
although a more extensive analysis of the robustness of this result (e.g., by varying the initial turbulent state, or instantaneously perturbing the saturated state with strong external fluctuations) could certainly help bridge between our numerical findings and the application to the ICM.

Nevertheless, what we see as the major impediment to this crucial task is the fact that to this date there is still no compelling evidence of the presence of the MTI in the periphery of galaxy clusters neither from cosmological simulations \citep[even without any sort of suppression of thermal conductivity][]{Ruszkowski2011,Kannan2017,Barnes2019} nor observations \citep[and it is likely that this will not be solved by the next generation of X-ray observatories such as Athena,][]{Kempf2023}.
While in the case of cosmological simulations this may be due to numerical limitations \citepalias[for a detailed discussion see][]{Perrone2022a}, the straight-up adoption of a subgrid closure as in Eq.~\eqref{eq:whistler_suppression_approx} clearly would not help address this problem. Granted, the suppression of thermal conduction is relevant to other physical processes in the ICM than just the MTI, such as the formation and survival of cold fronts \citep{Markevitch2007}, and the energy balance between cooling and heat transport \citep{Zakamska2003}. However, the consequences of the two main results of this work, the shutdown of the SSD and the beneficial effect of external turbulence on the MTI \citep[both very relevant to the ongoing debate on the source of relativistic electrons in the periphery of galaxy clusters; ][]{Cuciti2022,Beduzzi2023}, would be currently very difficult to assess within large-scale cosmological simulations. 
From a numerical perspective, a more fruitful avenue seems to progressively move toward  more realistic simulations of the ICM where the relevant MTI scales are adequately resolved, which might be easier to connect to observations. We do not rule out taking up this task ourselves in the future, or leave it as an open question for the community.

\section{Summary and conclusions}\label{sec:conclusions}

In this paper, which complements and expands our previous work \citepalias{Perrone2023}, we carried out an extensive investigation of the effects of whistler suppression of thermal conductivity on the MTI through small-scale simulations. We adopted a popular subgrid model for the thermal conductivity which smoothly interpolates between the Spitzer (at low plasma $\beta$) and the whistler-suppressed regime (at high $\beta$) and which is inspired by theoretical arguments \citep{Drake2021} and kinetic simulations \citep{Roberg-Clark2016,Roberg-Clark2018,Komarov2018}. 

Our model is characterized by a dimensionless parameter ($\alpha$) that encodes the information on the collisionality of a local patch of ICM plasma and the efficiency of thermal conduction on cluster-size scales. By varying $\alpha$ in the range that we can realistically expect in the periphery of galaxy clusters, we were able to study the consequences of weaker (low $\alpha$) or stronger suppression (large $\alpha$) of thermal conductivity on the MTI, which relies on effective conduction of heat along magnetic field lines, and that drives turbulence proportional to the thermal diffusivity. 

We found that for relatively weak suppression below the Spitzer value, MTI turbulence with whistler suppression appears phenomenologically rather similar to a simpler model of flat (i.e., spatially uniform) suppression, as the turbulent energy levels decrease proportionally to the volume-averaged suppressed thermal diffusivity \citep[][]{Perrone2022a}. In a similar fashion, the turbulent power spectra also resemble their counterparts with flat suppression.
This is in spite of the fact that whistler suppression is highly inhomogeneous in space,
with efficient thermal conductivity along thin bundles of strong magnetic fields (the Autobahns of the heat-carrying electrons, where magnetic field lines remain effectively isothermal), and local out-of-isothermality regions with high plasma $\beta$ (weak magnetic fields).
However, the similarities end when the suppression parameter $\alpha$ nears a critical value (which depends on the magnetic Reynolds number of the plasma), beyond which the MTI enters a new regime, the death-spiral, and the turbulence abruptly dies. This critical transition comes about via the suppression of the MTI-driven SSD, and represents one of the main results of this work. 
From simple estimates, we estimate that in the hot ICM the suppression parameter $\alpha$ can be on the order of the critical value, which, together with the fact that the physical conditions in the periphery of galaxy clusters can vary quite significantly in space and time, makes it a challenging regime to simulate. Therefore, modeling whistler suppression via a flat suppression of thermal conductivity should be done with care.

This unique connection between MTI turbulence and whistler suppression can be severed in the case where magnetic fields are propped up by some other physical process than the MTI. This leads us to the second main result of this study, which is that externally driven turbulence (e.g., representing turbulence driven by cluster mergers) can in  fact be  beneficial for the MTI with whistler suppression. The external turbulence sets a minimum value of thermal diffusivity due to its own SSD, in addition to which the MTI can then add its contribution. In this way, simulations where the MTI was previously dying can be be revived, and the resulting turbulent levels are greater than in analogous forcing-only runs. This shows that under certain circumstances the interaction between two sources of turbulence can be constructive and not necessarily detrimental.


Despite earlier assessments of its inhibition due to competing external turbulence and suppression of thermal conductivity by microinstabilities, the MTI has proven to be remarkably resilient, and in this work and in its companion paper \citepalias{Perrone2023} we have shown that one problem can alleviate the other. This clearly shows that the MTI is a complex dynamical process that requires detailed modeling beyond linear theory in order to draw firm conclusions. It should also serve as a cautionary tale to carefully consider the consequences of adopting certain subgrid models, and what new physics one might introduce in the problem by doing so.
Future work is clearly needed on two fronts: on   small scales, with the aim of strengthening our understanding of the impact of the whistler and other microinstabilities on the transport of electrons, and on   large scales, focusing on thorough studies of the MTI in a cosmological context.

\section*{Acknowledgements}
We would like to thank the anonymous referee for their constructive feedback, which led to a significant improvement in the presentation of the results of this work. The authors acknowledge support by the European Research Council under ERC-AdG grant PICOGAL-101019746. TB gratefully acknowledges funding from the European Union’s Horizon Europe research and innovation programme under the Marie Skłodowska-Curie grant agreement No 101106080.
LMP thanks Henrik Latter for helpful and generous advice on a draft version of the manuscript. 
Some of the plots in this work make use of the colormaps in the CMasher package \citep[][]{vanderVelden2020}.

%
The data underlying this article will be shared on reasonable request to the corresponding author. The numerical simulations produced for this work and in its companion paper \citepalias{Perrone2023} required approximately $2.4 \times 10^5$ CPU-core hours and $1.8 \pm 3 \, \si{MWh}$ of electrical power to run, emitting about $(0.7 \pm 0.1) \times 10^3 \, \si{kgCO_2e}$ based on the average conversion factor for the German power grid\footnote{Approximately $385 \, \si{gCO_2e/kWh}$ (\url{https://ourworldindata.org/grapher/carbon-intensity-electricity})}. This is more than the yearly $\mathrm{CO}_2$ emissions per capita related to energy use (household consumption, personal transportation, and consumption of goods and services) of any of the poorest 2 billion people in the world\footnote{\url{https://www.iea.org/commentaries/the-world-s-top-1-of-emitters-produce-over-1000-times-more-co2-than-the-bottom-1}}. 



\bibliographystyle{aa}
\bibliography{MTI-Whistler-LPaper}





\appendix

\section{Whistler-suppressed diffusivity in \textsc{SNOOPY}}\label{app:chi_derivation}

In this Appendix we show how to rewrite the analytical closure for whistler-suppressed diffusivity in terms of the Boussinesq variables. Consistent with the MHD approximation, we consider a two-species, quasi-neutral plasma ($n_\mathrm{i} \simeq n_\mathrm{e}$) with equal ion and electron temperatures ($T_\mathrm{i} \simeq T_\mathrm{e} \simeq T$).

The temperature gradient parallel to the magnetic field $L_{T,\parallel}$ in Eq.~\eqref{eq:whistler_suppression_approx} can be split into two terms: the contributions coming from the background profile ($\mathrm{d} T / \mathrm{d} z |_0$) plus the contribution of the local temperature perturbation ($\bm b \bcdot \bnabla \delta T$) as follows:
\begin{align}
    L_{T,\parallel}^{-1} &\equiv  \left\lvert\frac{\bm b \bcdot \bnabla T}{T} \right\rvert \simeq  \left\lvert b_z \left( \frac{\mathrm{d} \ln T}{\mathrm{d} z} \right)_0  + (\bm b \bcdot \bnabla )\frac{\delta T}{T_0} \right\rvert         \nonumber \\
    &= \frac{\omega_{\rmn{T}}^2}{g_0} \left\lvert  b_z  + \omega_{\rmn{T}}^{-2} \bm b \bcdot \bnabla \theta \right\rvert, \label{eq:parallel_tempscale}
\end{align}
where we have used the definition of the local MTI frequency in Eq.~\eqref{eq:freq} together with the definition of the buoyancy variable $\theta$ (Eq.~\ref{eq:theta_def}), and where $g_0$ is the local gravitational acceleration. We note that in writing Eq.~\eqref{eq:parallel_tempscale} we have neglected terms that are higher-order than $\bigO (\delta T / T_0)$, consistently with the Boussinesq approximation. Further, we assume hydrostatic equilibrium and write the pressure scale-height as $H = c_{\rmn{s}}^2 / g_0$, where $c_{\rmn{s}}^2 = \gamma T / m_\rmn{p}$ is the adiabatic sound speed for an hydrogenic plasma ($m_\rmn{p}$ is the proton mass). With these definitions, Eq.~\eqref{eq:whistler_suppression_approx} can be rewritten as follows:
\begin{align}
        \chi = \dfrac{\chi_0}{1 + \dfrac{1}{3} \dfrac{1}{\gamma} \left( \dfrac{\lambda_{\text{mfp,e}}}{H}\right) \left( \dfrac{H^2}{\chi_0/\omega_{\rmn{T}}}\right)  \left( \dfrac{2 \chi_0 \omega_{\rmn{T}}}{\varv_{\rmn{A}}^2 }\right)  \left\lvert b_z + \dfrac{ \bm b \bcdot \bnabla \theta}{\omega_{\rmn{T}}^2} \right\rvert} \label{eq:appendix-whistler-chi}
\end{align}
where we wrote the (electron) plasma beta as $\beta = 2 c_{\rmn{s}}^2 / (\gamma \varv_{\rmn{A}}^2)$. Replacing the definitions for the modified plasma $\tilde{\beta}$ (Eq.~\ref{eq:beta_tilde_def}), the suppression parameter $\alpha$ (Eq.~\ref{eq:alpha_def}), and the isothermality parameter (Eq.~\ref{eq:isolength_def})  we then obtain Equation~\eqref{eq:whistler-diffusivity}.

\section{Interruption of MTI growth with whistler suppression}\label{app:linear_theory}

\begin{figure*}
        \centering
        \includegraphics[width=0.7\linewidth]{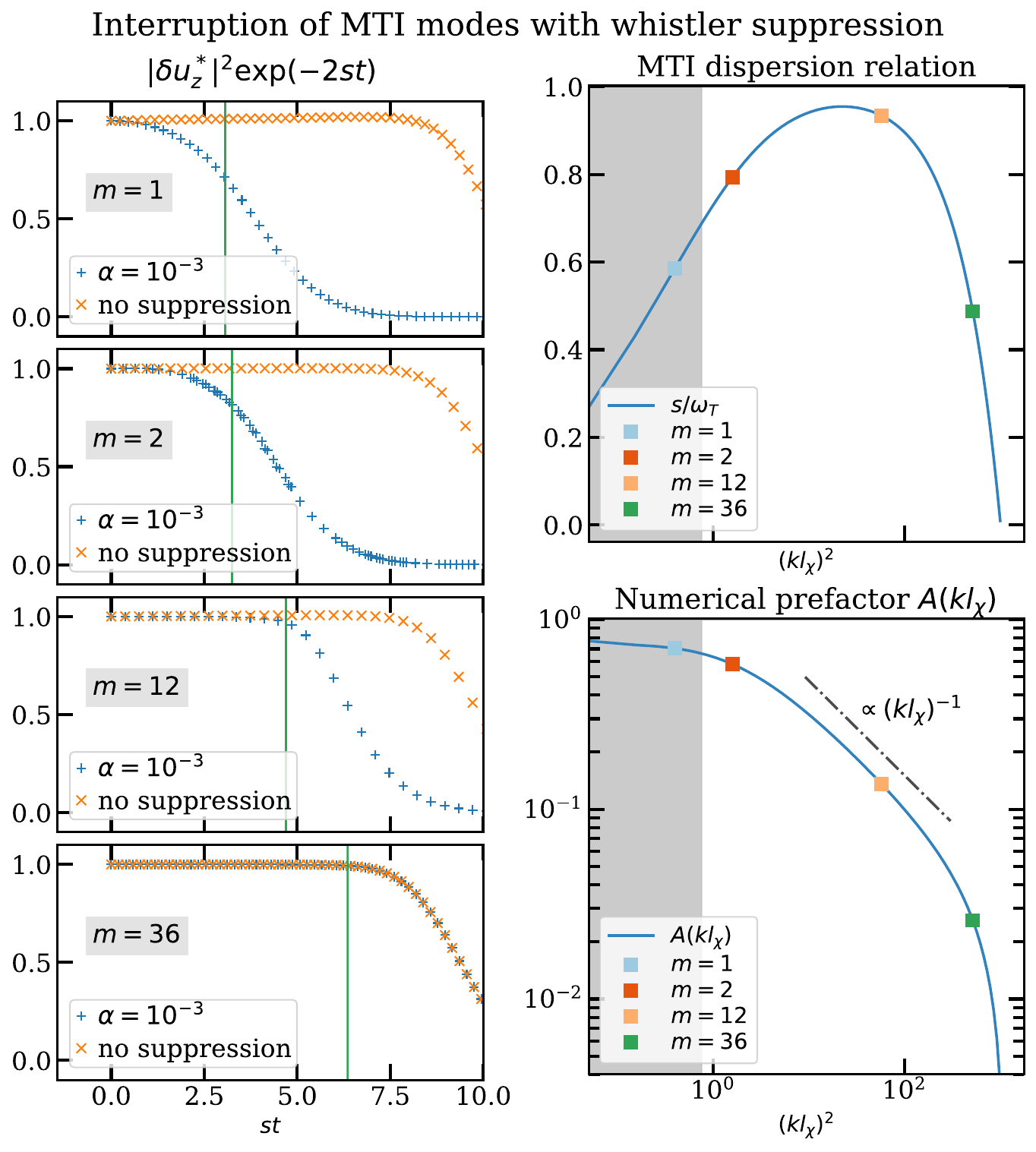}
        \caption{Interruption of exponential growth of MTI modes with whistler suppression of thermal diffusivity. Left column: Compensated amplitude of the MTI eigenmode as a function of time (nondimensionalized by the growth rate $s$) for perturbations of different wavenumbers: $k = 2 \pi m / L$, with $m=1,2,12,36$.  The green vertical lines show when the second term in Eq.~\eqref{eq:chi-first-order-eig} becomes unity, which roughly corresponds to the point where the thermal diffusivity is suppressed. Right column: Real part of MTI dispersion relation and growth rates for our runs (top panel), and amplitude of the numerical prefactor $A$ as a function of the wavenumber (bottom panel). The gray shaded area indicates the wavenumbers for which the growing MTI eigenmodes are in the regime of slow conduction (i.e., $ k^2  \chi_0/ s < 1)$.}
        \label{fig:mti_modes_interruption}
\end{figure*}

In this Appendix we show how the presence of whistler suppression of thermal diffusivity can interrupt the exponential growth of the MTI modes. As we discuss below, the effect of whistler suppression can be more or less severe, depending on the wave number of the MTI eigenmode, with modes growing on scales much smaller than the conduction length $l_\chi$ being the least affected.
We consider the onset of the MTI with whistler suppression from an initial state of equilibrium with $\bm u_0 = \mathbf{0}$, $\bm B_0 = B_0 \bm e_x$, $\theta_0 = 0$ (with constant gravity pointing in the $-\bm e_z$ direction), and looking at small perturbations $\delta \bm u^*, \delta \bm B^*, \delta \theta^* \sim \bigO (\epsilon)$, where $\epsilon$ is a small parameter. The starred quantities have been nondimensionalized using $l_{\chi,0} = (\chi_0 / \omega_{\rmn{T}})^{1/2}$ and $\omega_{\rmn{T}}$ as the units of length and time, respectively, as follows:
\begin{align}
        \delta \bm u^* \equiv \frac{\delta \bm u}{l_{\chi,0} \omega_{\rmn{T}}} , \hspace*{2em} \delta \bm B^* \equiv \frac{\delta \bm B}{B_0} , \hspace*{2em} \delta \theta^* \equiv \frac{\delta \theta}{l_{\chi,0}\omega_{\rmn{T}}^2}.
\end{align} 
For ease of notation, throughout this section we drop the ``0'' subscript in $l_{\chi,0}$, with the understanding that it always refers to the conduction length computed using the unsuppressed diffusivity $\chi_0$. At first order in $\epsilon$, Eq.~\eqref{eq:whistler-diffusivity} then reads:
\begin{align}
        \chi = \chi_0 \left[  1 - \frac{1}{3} \alpha \left(\frac{2 \chi_0 \omega_{\rmn{T}}}{B_0^2}\right) \left\lvert \delta B_z^* + \partial_x^* \theta^*  \right\rvert + \bigO (\epsilon^2) \right] , \label{eq:chi-first-order}
\end{align}
where $\partial_x^*$ is the partial derivative with respect to the nondimensionalized spatial variable variable $x^* \equiv x / l_\chi$. Because the magnetic field lines are initially isothermal, at zeroth order in $\epsilon$ the heat diffusivity is just $\chi_0$ and, consequently, the linearized buoyancy equation (Eq.~\ref{eq:buoyancy_eq}) is the same as that of the MTI without whistler suppression.  
As the MTI eigenmode grows exponentially, however, the magnetic field lines are progressively brought farther away from isothermality (quantified by the term proportional to $\left\lvert \delta B_z^* + \partial_x^* \theta^*  \right\rvert$ in Eq.~\ref{eq:chi-first-order}, which is the linearized version of $L_{T,\parallel}^{-1}$ in Eq.~\ref{eq:parallel_tempscale}), thus reducing $\chi$ and potentially thwarting the growth of the MTI. To study this effect, we follow the exponential evolution of an MTI eigenmode (we take the solution in \citetalias{Perrone2022} and choose $\bm k = k \bm e_x$ for simplicity) and plug it in Eq.~\eqref{eq:chi-first-order} to obtain
\begin{align}
        \chi \simeq \chi_0 \left[  1 - \frac{1}{3} \alpha \left(\frac{2 \chi_0 \omega_{\rmn{T}}}{B_0^2}\right) \delta u_{z,0}^* A (k l_{\chi}) | \sin (k l_{\chi} x^*) | \rmn{e}^{s t} \right],  \label{eq:chi-first-order-eig}
\end{align}
where $s$ is the growth rate of the eigenmode with wavenumber $k$, and $\delta u_{z,0}^*$ is the initial amplitude of the (nondimensionalized) vertical velocity. We note that in the previous equation we are allowed to replace the perturbed quantities with the exponentially growing MTI solution computed for the case without whistler suppression because, as we saw in Eq.~\eqref{eq:chi-first-order}, corrections to the eigenmode solution are higher order in $\epsilon$. In Eq.~\eqref{eq:chi-first-order-eig} $A (k l_{\chi})$ is a scale-dependent numerical prefactor whose definition is given by
\begin{align}
        A (k l_{\chi}) \equiv \frac{k l_\chi}{s/\omega_{\rmn{T}} + q (k l_\chi)^2}  \frac{ s/\omega_{\rmn{T}} + N^2/\omega_{\rmn{T}}^2 \left[ s/\omega_{\rmn{T}} + q (k l_\chi)^2\right]}{s/\omega_{\rmn{T}} + (k l_\chi)^2}, \label{eq:A-prefactor}
\end{align}
where $q = \eta / \chi_0$ is the Roberts number. The behavior of $A$ as a function of wavenumber is shown in Fig.~\ref{fig:mti_modes_interruption}, lower-right panel, where we see that its amplitude decreases as $kl_{\chi}$ increases. From a physical point of view, the magnitude of $A$ reflects the fact that for any given timescale (e.g., the MTI eigenmode $e$-folding time), anisotropic thermal diffusion can erase temperature gradients along magnetic field lines more efficiently if the scale of the perturbation is smaller than the corresponding conduction length $l_{\chi}$ (high $k l_{\chi}$).
Of course, irrespective of the wavenumber of the initial fluctuation, strict isothermality is  bound to be broken eventually as the MTI mode grows exponentially, but Fig.~\ref{fig:mti_modes_interruption} shows that if $kl_{\chi}$ is large then isothermality can be maintained for longer. We stress that these considerations are independent of whistler suppression and apply to the regular MTI with constant heat diffusivity as well. However, since our subgrid model of whistler-suppressed heat flux strongly depends on the level of isothermality along magnetic fields (given by the magnitude of $L_{T,\parallel}$), the choice of the initial scale of the perturbation will affect the onset of whistler suppression.

We verify this hypothesis numerically by simulating the exponential growth of an MTI eigenmode with whistler suppression and a weak horizontal magnetic field ($B_0 = 10^{-3} l_{\chi} \omega_{\rmn{T}}$, in Alfven velocity units): each run is initialized with a single MTI mode of increasing wavenumber $k = 2 \pi m / L$, where $m=1,2,12,36$. All runs have the same suppression parameter ($\alpha = 10^{-3}$), and initial amplitude of the MTI eigenmode ($\delta u_{z,0}^* = 10^{-4}$). The squared Brunt-V\"{a}is\"{a}l\"{a} frequency in units of the squared MTI frequency is $N^2/ \omega_{\rmn{T}}^2 = 0.1$, the unsuppressed diffusivity is $\chi_0 / (L^2 \omega_{\rmn{T}}) = 10^{-2}$, and we set low viscosity and low resistivity ($\nu = \eta = 10^{-3} \chi_0$). In the left column of Fig.~\ref{fig:mti_modes_interruption} we show the time evolution of the MTI eigenmode with whistler suppression (compensated by factoring out the exponential growth, blue ``$+$'') for each run, and compare it to an equivalent run without whistler suppression (orange crosses). As we can observe, in the runs with the lowest wavenumber the exponential growth of the initial perturbation ends after $1-2$ $e$-folding times (the times have been normalized by the growth rates of each run, so that the $x$ axis is in common). As we increase the wavenumber, however, the two curves lie on top of each other for longer, meaning that the suppression of thermal diffusion kicks in later and later, to the point that for $m=36$ we do not find any difference between the exponential growth phase of the run with and without whistler suppression.

\section{Numerical implementation of nonuniform thermal diffusivity in \textsc{SNOOPY}}\label{app:implementation_whistler_suppression}

\begin{figure*}
        \centering
        \includegraphics[width=0.9\linewidth]{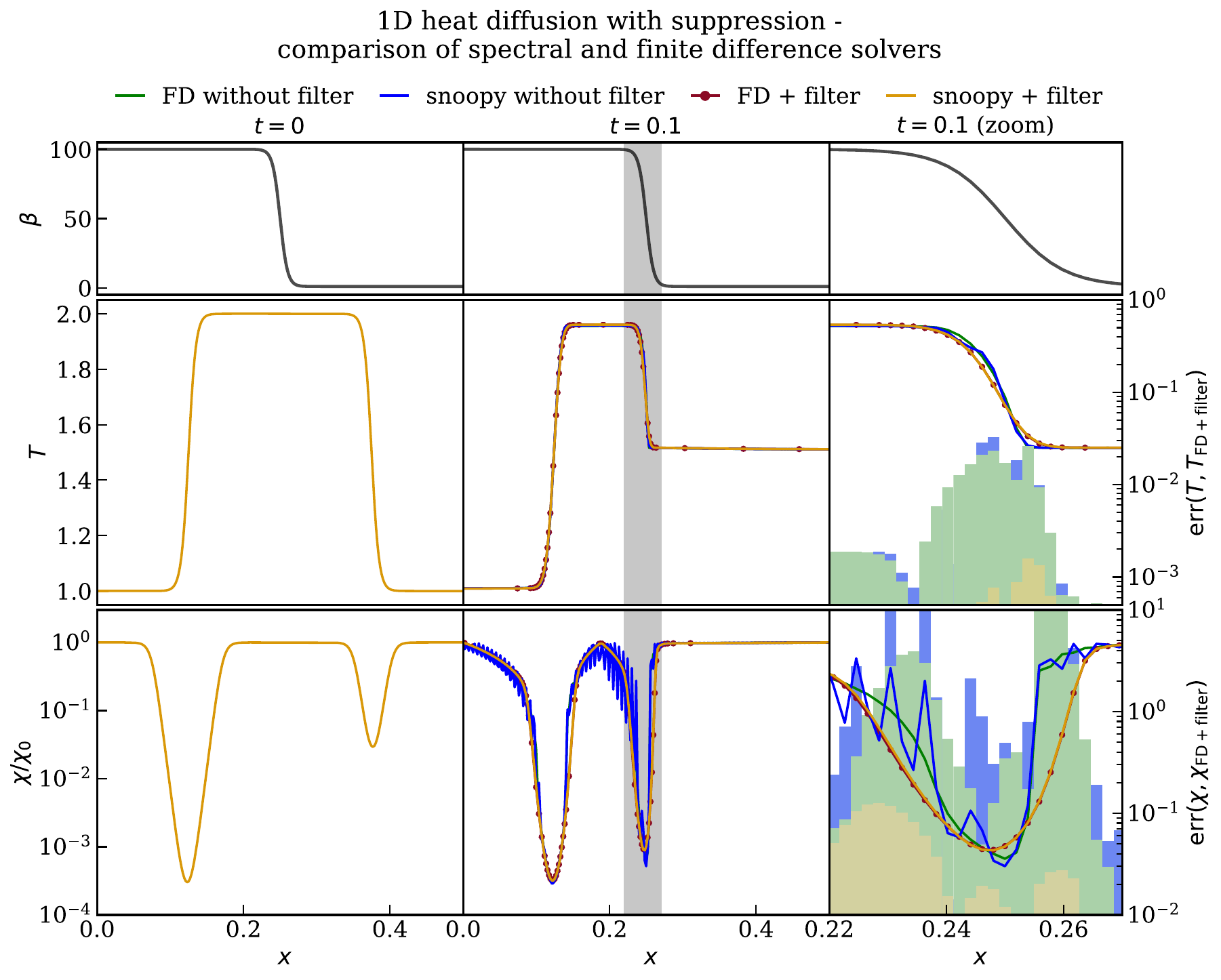}
        \caption{Solution of 1D heat diffusion equation with $\beta$-dependent suppression of thermal diffusivity (Eq.~\ref{eq:1d-model-diffusivity}). From left to right: Initial solution at time $t=0$, final solution at time $t=0.1$, and zoomed-in view of the final solution in the region $0.22 < x < 0.27$. The times are in units of $L^2 / \chi_0$. The different lines correspond to the solution obtained with a finite difference code without and with harmonic filtering (green and dark red, respectively), while the \textsc{Snoopy} solution is shown in blue (without filtering) and gold (with filtering). In the rightmost panel the bar plots show their relative error compared to the finite-difference solution with harmonic filter. The addition of harmonic filtering renders \textsc{Snoopy} more diffusive, but eliminates the problem of Gibbs oscillations at the grid-scale.}
        \label{fig:appendix-1d-comparison}
\end{figure*}

In this Appendix we describe the numerical implementation of whistler suppression in \textsc{Snoopy}. This is particularly important since pseudospectral codes such as \textsc{Snoopy} cannot accurately capture sharp discontinuities, which can realistically develop in the spatial profiles of $\chi$ given the high nonlinearity of Eq.~\eqref{eq:appendix-whistler-chi}. 

To showcase a typical scenario that we might encounter in our simulations, we plot in Fig.~\ref{fig:appendix-1d-comparison} the solution to a simple 1D periodic heat-diffusion problem with spatially dependent thermal diffusivity:
\begin{align}
    \frac{\partial }{\partial t} T(x,t) = \frac{\partial }{\partial x} \left( \chi (x) \frac{\partial T(x,t)}{\partial x}  \right).
\end{align}
For this test, we choose $\chi$ to depend on the plasma $\beta (x)$ (kept fixed during the simulation), and on the gradient of the temperature $T (x,t)$ in order to mimic the subgrid model of whistler suppression in Eq.~\eqref{eq:appendix-whistler-chi}:
\begin{align}
    \chi (x) = \frac{\chi_0}{1 + L \, \beta(x) \, \lvert \mathrm{d} \ln T (x,t) / \mathrm{d} x \rvert }, \label{eq:1d-model-diffusivity}
\end{align}
where $L=1$ is the size of the domain that extends over $[-0.5,0.5]$ (all length scales are given in code units). We initialize both $\beta (x)$ and $T(x,t=0)$ with smooth $\tanh$ profiles as follows:
\begin{align}
    \beta (x) &= 1 + \frac{\delta}{2} \left[ \tanh \left( \frac{x + 1/4}{a} \right) - \tanh \left( \frac{x - 1/4}{a} \right) \right] , \\
    T(x,t=0) &= 1 + \frac{1}{2}  \left[ \tanh \left( \frac{x + 3/8}{a} \right) - \tanh \left( \frac{x + 1/8}{a} \right) \right. \\
    & \left. + \tanh \left( \frac{x - 1/8}{a} \right) - \tanh \left( \frac{x - 3/8}{a} \right) \right],
\end{align}
with $\delta = 99$, $a = 0.01$, and compare the numerical solution obtained with \textsc{Snoopy} at time $t=0.1 L^2/\chi_0$ to that of a simple 1D solver
written in \textsc{Python} using a second-order finite difference spatial discretization and a fourth-order Runge-Kutta time-stepping (in Fig.~\ref{fig:appendix-1d-comparison} only the right-half of the periodic domain is shown). All the simulations have a resolution of $N=512$. The strong suppression of the thermal diffusivity for $x<0.25$ due to the high value of $\beta$ prevents the exchange of heat, preserving the initial temperature gradient. For $x>0.25$, on the other hand, the domain thermalizes rapidly, and the temperature profile becomes flat. This behavior is correctly reproduced with both types of numerical solvers. However, with \textsc{Snoopy} (solid blue line) we note the appearance of strong Gibbs oscillations around the finite-difference profile (solid green line) throughout the domain, and particularly in proximity to sharp gradients in $\chi$. We note that also the finite-difference solution appears to be slightly under-resolved, with low-amplitude oscillations around the ``real'' profile of $\chi$ near the grid scale (e.g., visible in the bottom right panel of Fig.~\ref{fig:appendix-1d-comparison} near $x\simeq 0.26$). For the finite-difference solution, we checked convergence increasing the resolution up to $N=2048$. 

To prevent the development of unphysical oscillations, and improve the numerical stability of \textsc{Snoopy}, we decided to implement a filter to the thermal diffusivity at each timestep. Taking inspiration from previous work on anisotropic diffusion \citep[e.g.,][]{Sharma2007} we opted for a simple 2-point centered harmonic averaging
\begin{align}
        \frac{1}{\bar{\chi}_i} = \frac{1}{2} \left( \frac{1}{\chi_{i+1}} + \frac{1}{\chi_{i-1}}\right),
\end{align}
where $\bar{\chi}$ is the filtered diffusivity, $i$ denotes the $i$-th cell, and $\chi_i = \chi (x_i)$. 
From a physical point of view, 
harmonic averaging of the thermal diffusivity corresponds to an arithmetic average of the heat diffusion times across two neighbor cells, and ensures that the slowest of the two rates prevails. This makes the code more diffusive, but is essential to prevent the formation of Gibbs oscillations with pseudospectral methods. 
The numerical implementation of an harmonic filter in \textsc{Snoopy} is very straightforward, since we can use again the property that 
harmonic averaging is just a regular arithmetic average of $1/\chi$, 
and we can recast the linear operator in real space into an equivalent local operator in Fourier space. For a generic function $\phi (x)$, with Fourier transform $\hat{\phi} (k)$ given by
\begin{align}
        \mathcal{F} \left[ \phi (x)\right] (k) \equiv  \hat{\phi} (k) = \sum_{i=0}^{N-1} \phi (x_i) e^{-i k x_i},
\end{align}
taking the Fourier transform of the arithmetic mean yields:
\begin{align}
 \mathcal{F} \left[ \frac{\phi (x + \Delta x) + \phi (x - \Delta x)}{2} \right] (k)
        = W(k) \hat{\phi} (k) \equiv \cos (k \Delta x) \hat{\phi} (k), \label{eq:harmonic_filter}
\end{align}
where we used the definition of the inverse transform
and the identity 
\begin{align}
        \frac{1}{N}\sum_{i=0}^{N-1}  \rmn{e}^{ -i (k-k') x_i} = \delta (k-k').
\end{align}
In Eq.~\eqref{eq:harmonic_filter}, $W(k)$ is the filter in Fourier space and $\Delta x$ the grid spacing. In higher dimensions ($d=2$ or $d=3$), Eq.~\eqref{eq:harmonic_filter} simply becomes:
\begin{align}
    W (\bm k) = \frac{1}{d} \sum_{i=1}^{d}  \cos (k_i \Delta h_i), 
\end{align}
where $\Delta h_i$ is the cell separation in the $i$-th direction. For a more detailed discussion on the use of filtering in pseudospectral codes (see, e.g., \citealt{Birdsall1991}). Using the result above, we implement the harmonic filter in \textsc{Snoopy} by computing $\chi^{-1}$ in real space, Fourier-transforming it, multiplying by $\cos (k \Delta x)$, transforming back and inverting to obtain $\bar{\chi}$. 

\begin{figure}
        \centering
        \includegraphics[width=1.0\linewidth]{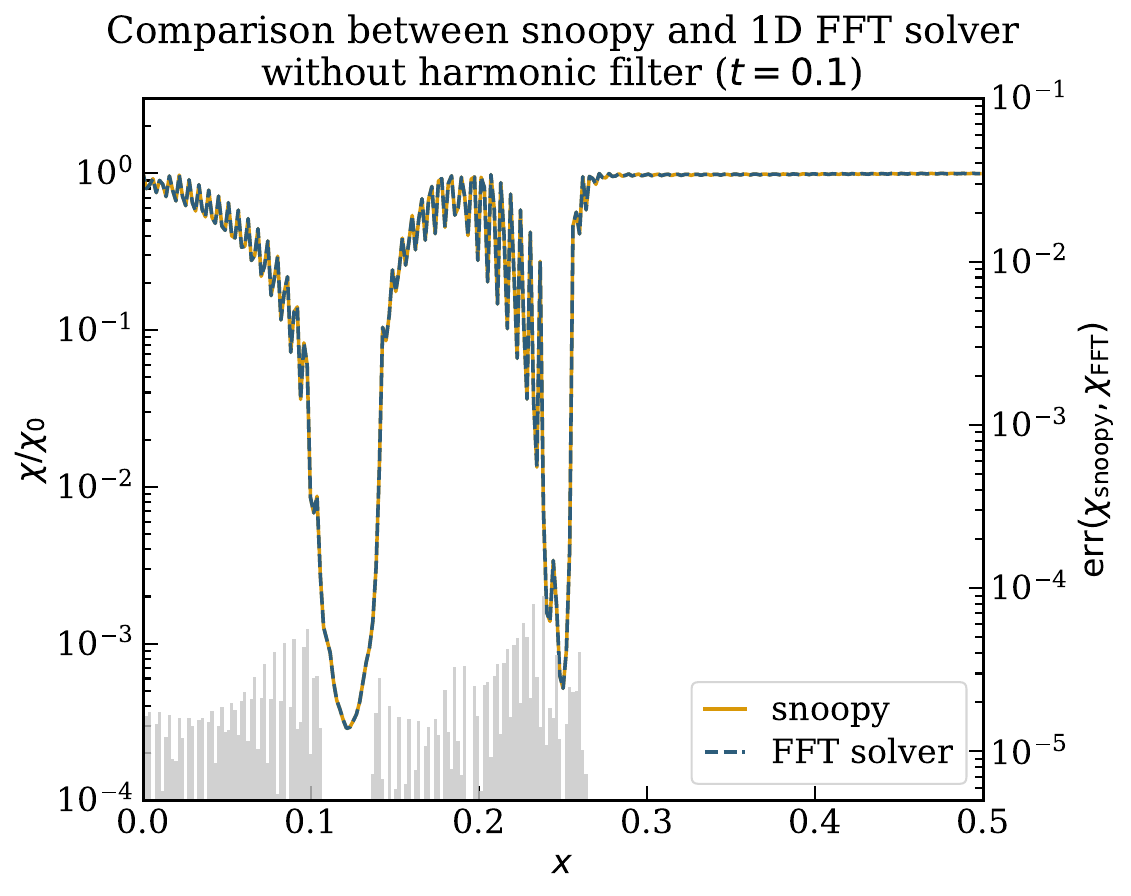}
        \caption{Comparison between the solutions to Eq.~\ref{eq:1d-model-diffusivity} obtained with \textsc{Snoopy} (solid gold line) and a 1D pseudospectral code written in \textsc{Python} (dashed blue line) without harmonic filtering. The development of Gibbs oscillations is not only a problem of \textsc{snoopy,} but a general feature of pseudospectral codes with strong gradients.}
        \label{fig:appendix-1d-snoopy-vs-fft}
\end{figure}

If we plot in Fig.~\ref{fig:appendix-1d-comparison} the finite-difference and the pseudospectral \textsc{Snoopy} solutions with harmonic averaging, we note that they are indeed more diffusive near strong temperature gradients, but we also see that now the pseudospectral solution does not show any sign of Gibbs oscillations and agrees very well with its finite-difference counterpart (relative error of less than $10\%$ in $\chi$ and less than $0.1\%$ in the temperature at the final time). We note that the presence of Gibbs oscillations in proximity to strong gradients is a generic feature of all pseudospectral codes and not just of \textsc{Snoopy}. This can be clearly seen in Fig.~\ref{fig:appendix-1d-snoopy-vs-fft} where we compare the solution obtained with \textsc{Snoopy} without harmonic filtering to that of a simple 1D FFT solver written in \textsc{Python} with fourth-order Runge-Kutta time-stepping and the same resolution: the two solutions agree up to a relative error of less than $10^{-4}$.

\begin{figure*}
        \centering
        \includegraphics[width=0.8\linewidth]{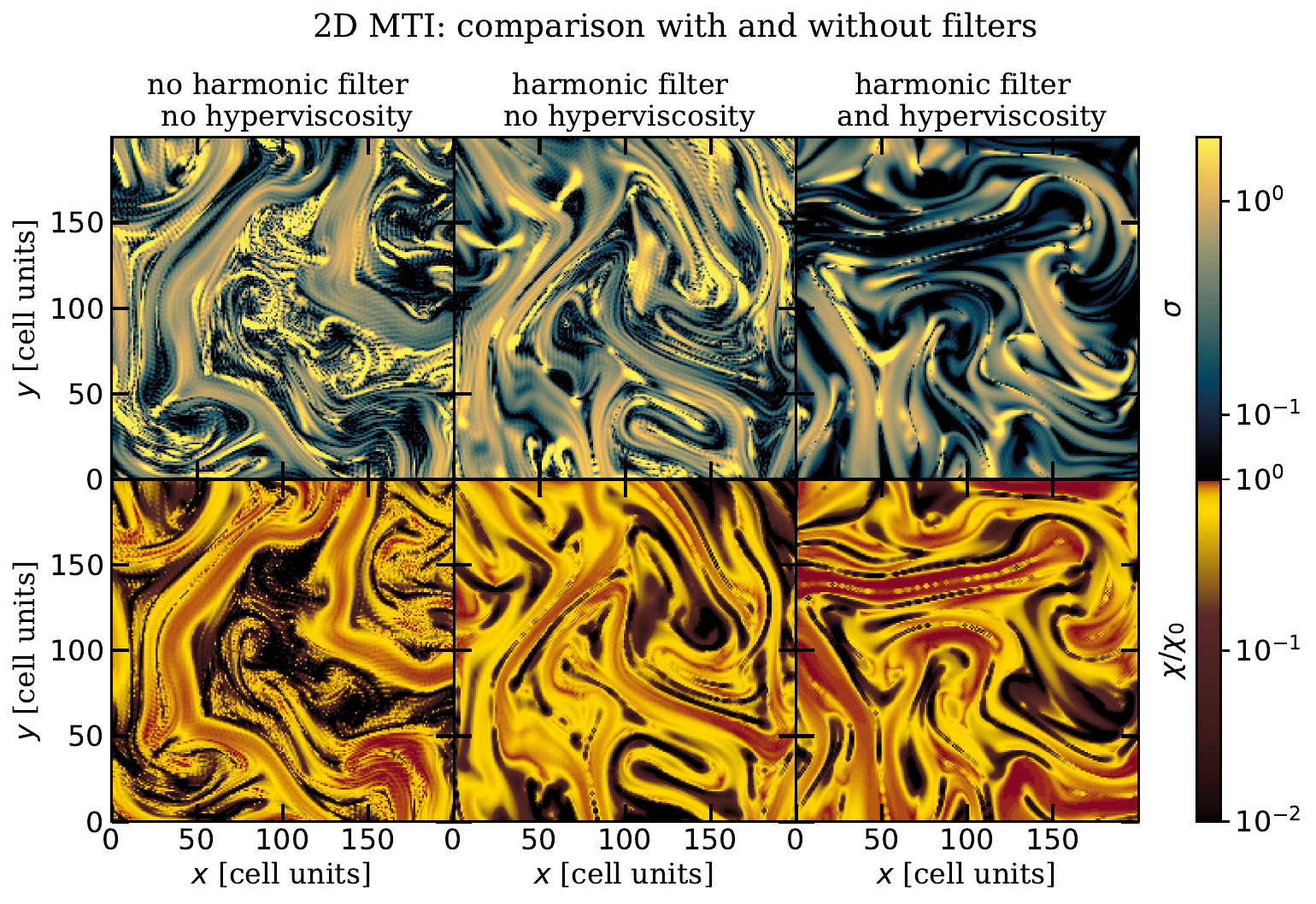}
        \caption{Comparison between different \textsc{Snoopy} runs of 2D MTI with whistler suppression and with different combinations of harmonic filtering of $\chi$ and hyperviscosity in $\theta$: without harmonic filtering and without hyperviscosity (left), with filtering and without hyperviscosity (middle), and with both (right). The parameters of the runs are otherwise identical and the same as the run with $\alpha = 4 \times 10^{-4}$. For visualization purposes we zoomed in on a region of size $200\times200$ cells (the resolution of the runs is $1024^2$). Only the run with both harmonic filtering and hyperviscosity shows no signs of Gibbs oscillations in either $\chi$ and the isothermality parameter $\sigma$, which depends on the gradient of $\theta$.}
        \label{fig:final_states_hyperdiffusion_filter_comparison}
\end{figure*}

To assess how the harmonic filter fares in a more realistic example, we show in Fig.~\ref{fig:final_states_hyperdiffusion_filter_comparison} a comparison between different versions of a 2D MTI simulation with whistler suppression ($\alpha = 4\times 10^{-4}$) without (left panel) and with harmonic averaging (middle panel). We note that for visualization purposes we zoom in on a $200\times200$-cell wide region (original resolution $1024^2$). Looking at the thermal diffusivity (lower row), we clearly find the presence of striations at the grid size throughout the domain, which are a sign of Gibbs oscillations similar to Fig.~\ref{fig:appendix-1d-comparison}; with the application of the harmonic filter these oscillations vanish. We note that while harmonic filtering is very effective at regularizing the thermal diffusivity, small-scale oscillations still persist in the isothermality variable $\sigma$, which is a function of the local temperature gradient. This is partly due to the anisotropic nature of thermal conductivity itself, which damps fluctuations along magnetic field lines but not across it, and partly due to the additional spatial dependence of $\chi$ in the anisotropic diffusion, which could lead to further numerical artifacts. As an additional safeguard against numerical instabilities, we decide to include a small amount of hyperdiffusivity (i.e., a higher-order diffusion operator) to the equation for $\theta$, the temperature fluctuation, in order to regularize the smallest scales in our simulation. We opted for a third-order ($\bnabla^6$) hyperdiffusion to allow for a wide scale separation between energy injection by the MTI and dissipation and checked that the resulting dynamics in spectral space are mostly unaffected (see \citetalias{Perrone2023} for further details). With the inclusion of hyperdiffusion we can see that no further oscillations are visible in our simulations (Fig.~\ref{fig:final_states_hyperdiffusion_filter_comparison} right column).

\end{document}